\documentclass[journal]{IEEEtran}

\usepackage{amsmath,amssymb,amsfonts}
\usepackage{bm}
\usepackage{graphicx}
\usepackage{stfloats}
\usepackage{booktabs}
\usepackage{cite}
\usepackage{balance}
\usepackage{algorithm}
\usepackage{algorithmic}
\usepackage{url}
\usepackage{placeins}
\usepackage[caption=false,font=footnotesize]{subfig}

\newcommand{\jj}{\mathrm{j}}


\begin{document}
	\bstctlcite{IEEEexample:BSTcontrol}
	
	\title{AFDM-ISAC With Fractional Delay–Doppler Coupling}
	
\author{Shaohua Li, Cunhua Pan, Hong Ren, Ruizhe Wang,
	and Jiangzhou Wang,~\IEEEmembership{Fellow,~IEEE}%
	\thanks{The authors are with National Mobile Communications Research
		Laboratory, Southeast University, Nanjing, China.
		(e-mail: 230268176, cpan, hren, rzw, j.z.wang@seu.edu.cn).
		\\ \hspace*{2em}\textit{Corresponding author: Cunhua Pan and Hong Ren.}}}

	\maketitle
	
	\begin{abstract}

			Affine frequency division multiplexing (AFDM) is a promising
			chirp-based multicarrier waveform for high-mobility integrated
			sensing and communication (ISAC). Accurate angle, delay, and
			Doppler estimation is essential for AFDM sensing. Since target
			delays and Doppler shifts are generally continuous-valued,
			representing them on a discrete delay--Doppler grid causes
			energy leakage and peak displacement in the discrete affine
			Fourier transform (DAFT) domain. The AFDM chirp also induces
			delay--Doppler coupling in the DAFT-domain response. The resulting DAFT-domain matching-score surface exhibits
			a local ridge that is not aligned with the normalized-delay
			and normalized-Doppler axes. To address these issues, this paper investigates joint estimation
			of angle and continuous-valued delay--Doppler parameters for a
			colocated AFDM-ISAC sensing architecture. A transform-domain sparse sensing
			model is formulated from the fractional DAFT-domain response.
			Based on this model, a coupled-coordinate Newtonized
			orthogonal matching pursuit (CC-NOMP) estimator is developed.
			CC-NOMP uses the AFDM-induced coupling coordinate to
			parameterize the dominant local ridge. It combines coupled-coordinate
			Newton refinement with safeguarded updates, coupling-aligned delay
			refinement, and cyclic multi-target refinement to estimate angle,
			continuous normalized delay, and normalized Doppler. A deterministic Cram\'er--Rao bound and a dominant-order complexity analysis are also derived. Simulation results with continuous-valued
			off-grid target parameters show that CC-NOMP achieves lower
			delay and Doppler error floors than the considered baselines
			while maintaining comparable angle-estimation accuracy.
 
	\end{abstract}
	
	
	\section{Introduction}
	
Integrated sensing and communication (ISAC) allows wireless
infrastructure to share spectrum, hardware, waveform, and
signal-processing resources for data transmission and environmental
sensing
\cite{Hassanien2019SPM,Liu2020TCOMM,Liu2022JSAC}.
Its fundamental limits, waveform designs, and mobile-network
architectures have also been studied extensively
\cite{ALiu2022COMST,Zhong2023JointWaveform,Ren2026TwoTimescale}.
This shared design is attractive for high-mobility wireless networks,
where communication channels and sensing echoes change rapidly over
time. For the sensing function, angle, delay, and Doppler estimation are
key tasks because they support localization, tracking, channel
awareness, and motion inference. Therefore, an effective estimator
should consider not only the array geometry but also the transmitted
waveform structure and the transform-domain observation used at the
receiver.

Delay--Doppler waveforms have been extensively studied for
communication and sensing over doubly selective channels
\cite{Hadani2017WCNC,Mohammed2022BITS,Gaudio2020TWC,
	Yuan2024MWC,Zhang2023OTFSRadarSensing}. Orthogonal
time-frequency space (OTFS) and affine frequency division
multiplexing (AFDM) have both been investigated for ISAC in
doubly dispersive channels \cite{Rou2024SPM}. While OTFS maps
information symbols to the delay--Doppler domain, AFDM employs
a chirp-domain discrete affine Fourier transform (DAFT) with two
tunable affine parameters, enabling adaptation to the channel
delay--Doppler spread and facilitating path separation in the
DAFT domain \cite{Rou2024SPM,Bemani2023TWC}. AFDM
channel-estimation studies and recent surveys have further
examined pilot design, implementation characteristics, and
high-mobility applications
\cite{Benzine2023GLOBECOM,Yin2025MWC,Rou2026AFDM6G}.
For sensing, the affine-parameter configuration shapes the
DAFT-domain target response, including the delay--Doppler
coupling considered in this paper. The ambiguity behavior,
pulse-shaping characteristics, and DAFT-domain sidelobe structure
also affect target-parameter estimation
\cite{Bedeer2025AFDMAmbiguity,Yin2026JSACAmbiguity,
	Ni2026PulseShapedAF}.

Recent AFDM-ISAC studies have examined several sensing architectures.
Matched-filtering and related time-domain or DAFT-domain processing
methods have been developed for range and Doppler estimation under
data-aided or pilot-aided settings
\cite{Ni2022ISWCS,Bemani2024WCL,Ni2025TWC}. System-level studies have
investigated waveform-parameter tradeoffs and unambiguous
delay--Doppler acquisition \cite{Ni2025TWC}, while index-modulated
AFDM sensing waveforms \cite{Temiz2025SPAWC} and pilot designs guided by the Cram\'er--Rao bound (CRB) and
ambiguity function \cite{Zhang2025AFDMISACFramework} have also been considered. Bistatic
AFDM sensing has been studied in static-scatterer environments
\cite{Zhu2024WCL}. Reconstruction, subspace processing, and
learning-enabled multi-target super-resolution sensing have also been
investigated
\cite{Luo2025Reconstruction,Chen2025MUSIC,Li2026DeepLearningAFDM}.
Other studies have addressed DAFT-domain interference cancellation,
multipath power-delay-profile estimation, joint channel--data--radar
parameter estimation, and blind bistatic parameter estimation
\cite{Lu2025DAFTIC,Xiao2026TCOMM,Ranasinghe2025TWC,
	Ranasinghe2025Blind}. These studies collectively demonstrated the practical value of
	AFDM-based sensing. However, relatively limited attention has been devoted to
	multi-target estimation with fractional delay--Doppler parameters.

Related AFDM sensing studies have also considered parameter acquisition
beyond integer or strictly grid-aligned models. For mixed near-field
and far-field sensing, a tensor-decomposition-based
angle--delay--Doppler estimator constructs a structured canonical
polyadic tensor and exploits factor separability to recover angle
together with fractional delay and Doppler parameters
\cite{Luo2025IoTJ}. A low-complexity fractional delay--Doppler
feature-extraction scheme characterizes the scatter and leakage patterns
induced by fractional parameters in the DAFT domain and estimates target
range and velocity under an embedded-pilot receiver
\cite{Zhu2026WCL}. An off-grid sparse Bayesian learning method
incorporates Doppler-grid offsets into the AFDM sensing model for joint
range--velocity estimation \cite{Luo2025SBL}. More recently, a
frequency-modulated continuous-wave (FMCW)-based AFDM formulation established a delay--Doppler-parameterized
DAFT-domain model and developed matched-filtering processors that
explicitly account for chirp-induced delay--Doppler coupling
\cite{Zhu2026FMCWAFDM}. These studies addressed fractional delay--Doppler effects, off-grid
Doppler estimation, and coupling-aware processing under different
signal models and receiver structures. Nevertheless, accurate
multi-target estimation of continuous delay and Doppler parameters
from multiple known data-bearing AFDM symbols remains challenging
because of off-grid leakage, AFDM-induced delay--Doppler coupling,
and overlapping target responses.

This challenge arises because the chirp-dependent DAFT-domain
response couples delay and Doppler within the local response of each
target. As a result, the dominant local variation follows a ridge-like
direction rather than two independent physical-coordinate directions.
Direct refinement in the physical delay--Doppler coordinates
does not explicitly expose this ridge and may therefore be
sensitive to strong local delay--Doppler coupling, especially
when multiple target responses overlap in the DAFT domain.
These observations call for an estimator that performs continuous
parameter refinement while accounting for the coupling structure of
the AFDM response.

Sparse recovery provides a suitable starting point for this problem.
More generally, basis mismatch can spread a continuously parameterized
component over several discrete dictionary atoms and degrade sparse
support recovery \cite{Chi2011BasisMismatch}. Orthogonal matching
pursuit (OMP) provides greedy coarse-support detection
\cite{Tropp2007TIT}, whereas Newtonized OMP (NOMP) combines sparse
component detection with Newton-based continuous refinement
\cite{Ramasamy2016TSP,Shah2025NOMPOFDM}. However, a direct application of NOMP to the adopted AFDM
sensing model refines delay and Doppler directly in the physical
coordinates and does not explicitly expose the dominant local
ridge direction. In the adopted DAFT-domain model, the
auxiliary coupling coordinate provides a ridge-aligned
parameterization for safeguarded updating, Doppler
reconstruction, and the subsequent coupling-aligned delay
refinement. This motivates a coupling-aware refinement strategy
for AFDM-based fractional delay--Doppler estimation.

Accordingly, this paper develops a coupled-coordinate NOMP estimator
for AFDM sensing. The main contributions are summarized as follows.
	
	\begin{itemize}
		\item A transform-domain sparse sensing model is formulated for the considered AFDM-ISAC sensing problem with continuous normalized delay and normalized Doppler shifts. The model describes the fractional delay--Doppler response in the DAFT domain, including off-grid phase variation and energy leakage.

		\item The AFDM chirp-induced delay--Doppler coupling is represented
		through an auxiliary DAFT-domain coordinate. This coordinate
		characterizes the local ridge structure and supports safeguarded
		coupled-coordinate Newton refinement and Doppler reconstruction.
		
		\item A coupled-coordinate Newtonized orthogonal matching pursuit
		(CC-NOMP) estimator is developed for joint angle, normalized
		delay, and normalized Doppler estimation. It combines coarse
		support detection, coupled-coordinate Newton refinement with
		safeguarded updates, local physical-coordinate refinement,
		coupling-aligned delay refinement, coefficient re-estimation,
		and cyclic multi-target refinement. Coupling-aligned delay
		refinement improves physical delay--Doppler estimation,
		while cyclic multi-target refinement mitigates residual inter-target
		interference.
		
		\item A deterministic CRB benchmark and a
		dominant-order complexity analysis are provided. The CRB treats
		the reflection coefficients as nuisance parameters, and the complexity
		analysis identifies the dominant computational costs of CC-NOMP and
		the considered baselines.
	\end{itemize}
	
	The remainder of this paper is organized as follows. Section~II presents the system model. Section~III develops the proposed estimation framework. Section~IV provides the performance analysis, including the CRB benchmark and complexity analysis. Section~V presents the numerical results. Section~VI concludes this paper.
	
	\section{System Model}
	\label{sec:system}
	
	\begin{figure}[!t]
		\centering
		\includegraphics[width=\columnwidth]{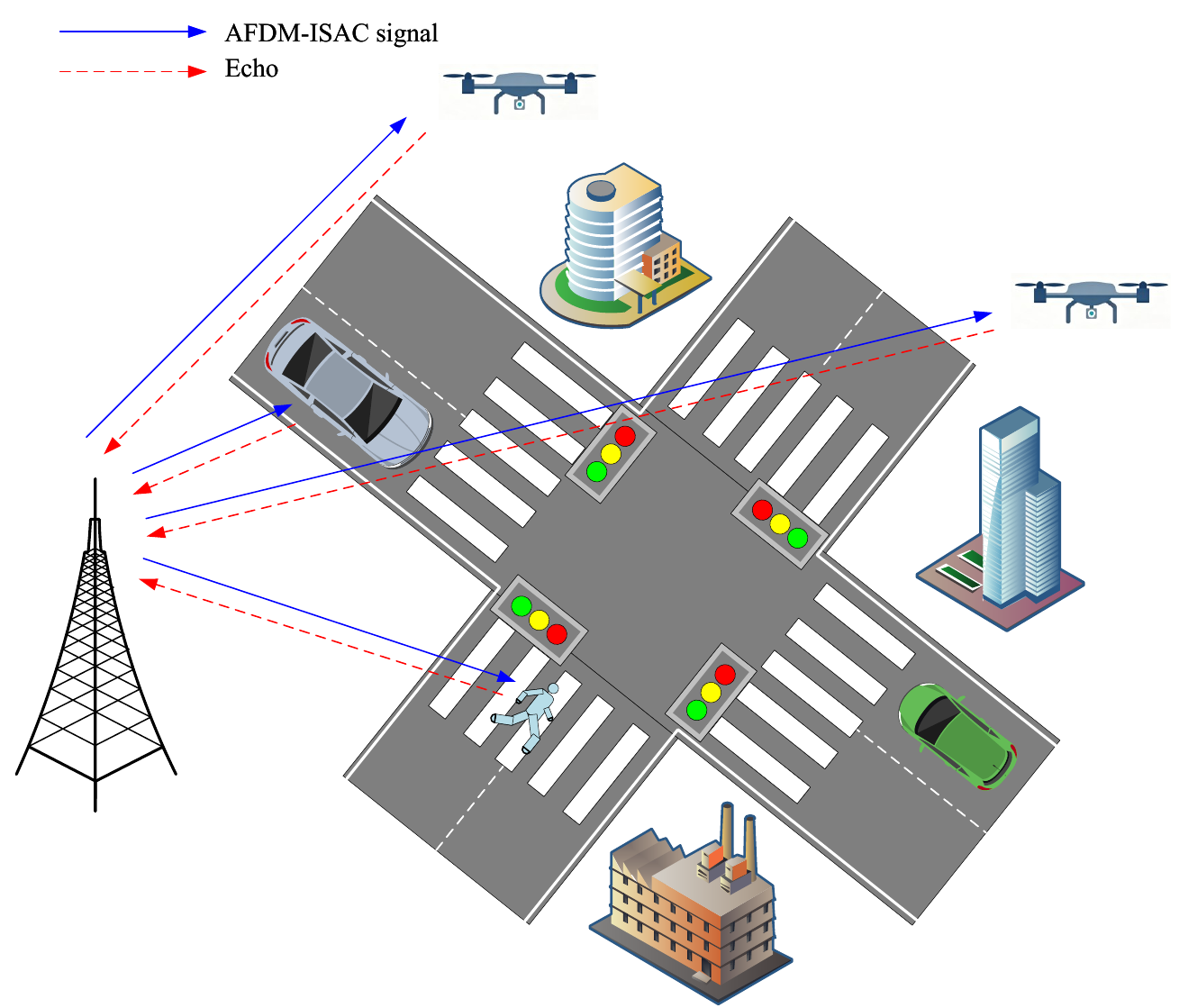}
		\caption{Colocated AFDM-ISAC sensing scenario.}
		\label{fig:system_model}
	\end{figure}
	
	Consider a colocated AFDM-ISAC sensing scenario comprising
	a sensing node and $P$ far-field point targets, as illustrated
	in Fig.~1. The sensing node employs a transmitter
	with a single antenna and a receive array with
	$N_{\mathrm R}$ elements. The transmitter radiates an AFDM
	communication waveform carrying data, which also illuminates
	the targets, while the receive array collects the reflected
	echoes for sensing. It is assumed that the transmit antenna and receive array use
	physically separated antenna apertures and that direct leakage
	from the transmit antenna to the receive array is negligible.
	The separation between the transmit and receive apertures is
	assumed to be negligible compared with the target ranges.
	Accordingly, a common propagation geometry is adopted under
	the far field assumption. The
	receive array is a uniform linear array with spacing
	$d_{\mathrm R}=\lambda/2$, where $\lambda$ denotes the carrier
	wavelength. The target number $P$ is assumed to be known.
	
	Let $T_{\mathrm{s}}$ denote the receiver sampling interval, with sampling frequency $f_{\mathrm{s}}=1/T_{\mathrm{s}}=B$. Each AFDM symbol contains $N_{\mathrm c}$ subcarriers, with subcarrier spacing $\Delta f=1/(N_{\mathrm c}T_{\mathrm{s}})$ and useful duration $T=N_{\mathrm c}T_{\mathrm{s}}$. Each transmission block consists of $N$ AFDM symbols, indexed by $n=1,\cdots,N$. Let $x_{n,m}$ denote the quadrature amplitude modulation (QAM) symbol transmitted in the $n$th AFDM symbol on DAFT-domain subcarrier $m$. In the considered colocated data-aided sensing architecture,
	the transmitted symbols $x_{n,m}$ are assumed to be
	available at the sensing receiver. A chirp-periodic prefix (CPP) of length $L_{\mathrm{CPP}}$ is appended to each AFDM symbol\cite{Bemani2023TWC}. The time-domain sample is given by
	\begin{equation}
		s_n(k)=\frac{1}{\sqrt{N_{\mathrm c}}}
		\sum_{m=0}^{N_{\mathrm c}-1}
		x_{n,m}
		e^{\jj2\pi(c_1 \xi_n^2(k)+m\xi_n(k)/N_{\mathrm c}+c_2 m^2)},
		\label{eq:tx_signal}
	\end{equation}
	where $c_1$ and $c_2$ are the AFDM chirp parameters. The within-symbol time index $\xi_n(k)$ is defined as
	\begin{equation}
		\xi_n(k)
		= k-L_{\mathrm{CPP}}-(n-1)(N_{\mathrm c}+L_{\mathrm{CPP}}).
		\label{eq:local_time_index}
	\end{equation}
	The sample indices of the $n$th AFDM symbol are
	\begin{equation}
		k=(n-1)(N_{\mathrm c}+L_{\mathrm{CPP}}),\cdots,n(N_{\mathrm c}+L_{\mathrm{CPP}})-1 .
		\label{eq:tx_sample_index}
	\end{equation}
	
	After CPP removal, the useful sample observed at receive antenna $n_{\mathrm{R}}$ is modeled as
	\begin{align}
		y_{n,n_{\mathrm{R}}}(k)
		&=\sum_{p=1}^{P}
		\alpha_p a_{\mathrm{R}}(n_{\mathrm{R}},\theta_p)
		\nonumber\\
		&\quad \times
		s_n(k-\ell_p)
		e^{\jj2\pi\nu_pk/N_{\mathrm c}}
		+w_{n,n_{\mathrm{R}}}(k),
		\label{eq:rx_time}
	\end{align}
	where
	\begin{equation}
		k=(n-1)(N_{\mathrm c}+L_{\mathrm{CPP}})+L_{\mathrm{CPP}},\cdots,n(N_{\mathrm c}+L_{\mathrm{CPP}})-1 .
		\label{eq:rx_useful_index}
	\end{equation}
	For the $p$th target, $\alpha_p$, $\theta_p$, $\ell_p$, and $\nu_p$ denote the reflection coefficient, angle, normalized delay, and normalized Doppler shift, respectively. The receive steering factor is given by
	\begin{equation}
		a_{\mathrm{R}}(n_{\mathrm{R}},\theta)
		=e^{\jj2\pi d_{\mathrm{R}}(n_{\mathrm{R}}-1)\sin\theta/\lambda}.
		\label{eq:rx_steering_factor}
	\end{equation}
	
	The normalized delay and normalized Doppler shift are decomposed as
	\begin{equation}
		\ell_p=\frac{\tau_p}{T_{\mathrm{s}}}=l_p+\iota_p,\qquad
		\nu_p=f_{\mathrm{d},p}T=V_p+v_p,
		\label{eq:fractional_params}
	\end{equation}
	where $l_p\in\{0,\cdots,l_{\mathrm{max}}\}$ and $V_p\in\{-v_{\mathrm{max}},\cdots,v_{\mathrm{max}}\}$ are integer grid indices, while $\iota_p\in(-1/2,1/2]$ and $v_p\in(-1/2,1/2]$ denote the corresponding fractional offsets. The bounds $l_{\mathrm{max}}$ and $v_{\mathrm{max}}$ are assumed to be known. The CPP length is selected to cover the maximum target delay.
	
	Since $\ell_p$ may be fractional, the shifted waveform $s_n(k-\ell_p)$
	is modeled through the adopted bandlimited circular fractional-delay
	representation, in line with fractional delay--Doppler modeling in
	AFDM sensing\cite{Zhu2026WCL}, which leads to the DAFT-domain phase response
	derived below. The same representation and DAFT normalization are used in both data generation and dictionary construction. The noise samples $w_{n,n_{\mathrm{R}}}(k)$ are modeled as independent complex Gaussian random variables with variance $\sigma_w^2$. Define
	\begin{equation}
		t_n=(n-1)(N_{\mathrm c}+L_{\mathrm{CPP}})+L_{\mathrm{CPP}},\qquad
		k'=k-t_n,
		\label{eq:local_index_def}
	\end{equation}
	where $k'=0,\cdots,N_{\mathrm c}-1$. For compact notation, define
	\begin{equation}
		\Gamma_n(\nu)=e^{\jj2\pi\nu t_n/N_{\mathrm c}}.
		\label{eq:gamma_def}
	\end{equation}
	Using the DAFT-domain periodicity of AFDM, \eqref{eq:rx_time} can be rewritten with the within-symbol sample index as
	\begin{align}
		y_{n,n_{\mathrm{R}}}(k')
		&=\sum_{p=1}^{P}\alpha_p a_{\mathrm{R}}(n_{\mathrm{R}},\theta_p)\Gamma_n(\nu_p)
		\nonumber\\
		&\quad \times
		\frac{1}{\sqrt{N_{\mathrm c}}}\sum_{m=0}^{N_{\mathrm c}-1}
		x_{n,m}
		\nonumber\\
		&\quad \times
		e^{\jj2\pi (c_1(k'-\ell_p)^2+m(k'-\ell_p)/N_{\mathrm c})}
		\nonumber\\
		&\quad \times
		e^{\jj2\pi c_2m^2}
		e^{\jj2\pi\nu_pk'/N_{\mathrm c}}
		+w_{n,n_{\mathrm{R}}}(k').
		\label{eq:rx_periodic}
	\end{align}
	
	After DAFT demodulation, the received signal at DAFT-domain bin $m'$ is obtained as
	\begin{align}
		y_{n,n_{\mathrm{R}}}(m')
		&=\sum_{p=1}^{P}\alpha_p a_{\mathrm{R}}(n_{\mathrm{R}},\theta_p)\Gamma_n(\nu_p)
		\nonumber\\
		&\quad \times
		\frac{1}{N_{\mathrm c}}\sum_{m=0}^{N_{\mathrm c}-1}
		x_{n,m}
		\nonumber\\
		&\quad \times
		e^{\jj2\pi(c_1\ell_p^2-m\ell_p/N_{\mathrm c})}
		e^{\jj2\pi c_2(m^2-m'^2)}
		\nonumber\\
		&\quad \times
		\frac{e^{\jj2\pi\zeta_{p,m,m'}}-1}
		{e^{\jj(2\pi/N_{\mathrm c})\zeta_{p,m,m'}}-1}
		+w_{n,n_{\mathrm{R}}}(m'),
		\label{eq:daft_response}
	\end{align}
	where $m'=0,\cdots,N_{\mathrm c}-1$. Since the adopted DAFT normalization is unitary, the transformed noise samples remain complex Gaussian with variance $\sigma_w^2$. The last factor in (11) is the closed-form expression of the finite
	geometric sum
\begin{align}
	D_{N_{\mathrm c}}(\zeta)=\sum_{q=0}^{N_{\mathrm c}-1} e^{\jj2\pi q\zeta/N_{\mathrm c}}
	=\frac{e^{\jj2\pi\zeta}-1}{e^{\jj(2\pi/N_{\mathrm c})\zeta}-1}.
\end{align}
	When the denominator is zero, \(D_{N_{\mathrm c}}(\zeta)\) is evaluated by its
	limit. Equation~\eqref{eq:daft_response} follows by substituting \eqref{eq:rx_periodic} into the DAFT demodulation expression and evaluating the resulting finite geometric sum. The detailed derivation is provided in Appendix~A.
	
	With the coupling coefficient $\rho=2N_{\mathrm c} c_1$, the phase term that determines the DAFT-domain displacement is
	\begin{equation}
		\zeta_{p,m,m'}=(m-m')-\rho\ell_p+\nu_p .
		\label{eq:coupling_phase}
	\end{equation}
	Thus, the dominant displacement depends on the coupled delay--Doppler quantity rather than on delay and Doppler independently. This motivates the DAFT-domain coupling coordinate
	\begin{equation}
		\eta_p=\nu_p-\rho\ell_p .
		\label{eq:eta_def}
	\end{equation}
	Here, $\eta_p$ is an auxiliary DAFT-domain coupling coordinate, not an additional physical parameter or normalized Doppler shift. It provides a parameterization aligned with the AFDM chirp-induced ridge, and the normalized Doppler shift is recovered as $\nu_p=\eta_p+\rho\ell_p$. Delay and Doppler remain distinguishable through the residual delay-dependent phase terms, Doppler-dependent inter-symbol phase rotation, and accumulation across AFDM symbols.
	
	The demodulated samples over receive antennas, AFDM symbols, and DAFT-domain bins are stacked into $\mathbf{y}\in\mathbb{C}^{N_{\mathrm{R}}NN_{\mathrm c}\times 1}$ as
	\begin{equation}
		\relax[\mathbf{y}]_{(n_{\mathrm{R}}-1)NN_{\mathrm c}+(n-1)N_{\mathrm c}+m'+1}=y_{n,n_{\mathrm{R}}}(m').
		\label{eq:stack_index}
	\end{equation}
	The stacked sensing model is written as
	\begin{equation}
		\mathbf{y}=\sum_{p=1}^{P}\alpha_p\mathbf{a}(\theta_p,\ell_p,\nu_p)+\mathbf{w},
		\label{eq:stacked_model}
	\end{equation}
	where $\mathbf{w}\sim\mathcal{CN}(\mathbf{0},\sigma_w^2\mathbf{I}_{N_{\mathrm{R}}NN_{\mathrm c}})$. For generic arguments, define
	\begin{equation}
		\begin{aligned}
			\psi_{m,m'}(\ell,\nu)
			&=
			e^{\jj2\pi\left[c_1\ell^2-m\ell/N_{\mathrm c}+c_2(m^2-m'^2)\right]}
			\\
			&\quad \times
			\frac{e^{\jj2\pi\zeta_{m,m'}(\ell,\nu)}-1}
			{e^{\jj(2\pi/N_{\mathrm c})\zeta_{m,m'}(\ell,\nu)}-1},
		\end{aligned}
		\label{eq:freq_afdm_kernel}
	\end{equation}
	where $\zeta_{m,m'}(\ell,\nu)$ is obtained from \eqref{eq:coupling_phase} by replacing $(\ell_p,\nu_p)$ with $(\ell,\nu)$. The kernel $\psi_{m,m'}(\ell,\nu)$ characterizes the DAFT-domain delay--Doppler response from transmitted DAFT-domain subcarrier $m$ to received bin $m'$. The $(n_{\mathrm{R}},n,m')$th entry of $\mathbf{a}(\theta,\ell,\nu)$ is
	{
		\setlength{\abovedisplayskip}{4pt}
		\setlength{\belowdisplayskip}{4pt}
		\setlength{\abovedisplayshortskip}{3pt}
		\setlength{\belowdisplayshortskip}{3pt}
		\begin{equation}
			\begin{aligned}
				\relax[\mathbf{a}(\theta,\ell,\nu)]_{n_{\mathrm{R}},n,m'}
				&=a_{\mathrm{R}}(n_{\mathrm{R}},\theta)\Gamma_n(\nu)
				\\
				&\quad \times
				\frac{1}{N_{\mathrm c}}\sum_{m=0}^{N_{\mathrm c}-1}
				x_{n,m}\psi_{m,m'}(\ell,\nu).
			\end{aligned}
			\label{eq:atom_explicit}
		\end{equation}
	}
	The atom $\mathbf{a}(\theta,\ell,\nu)$ includes the receive-array response,
	inter-symbol Doppler phase, known QAM symbols, and fractional
	DAFT-domain delay--Doppler response. For off-grid $\ell$ and $\nu$,
	the finite-sum ratio in $\psi_{m,m'}(\ell,\nu)$ spreads the target
	response over multiple DAFT-domain bins. The dominant displacement
	is governed by $\nu-\rho\ell$, which yields a local ridge
	parameterized by $\eta=\nu-\rho\ell$, while the remaining phase
	terms preserve the information needed to recover $(\theta,\ell,\nu)$.
	
	\begin{figure}[t]
		\centering
		\includegraphics[width=0.95\columnwidth]{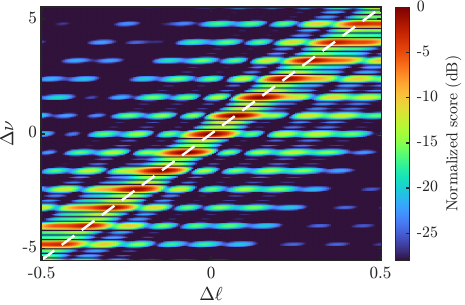}
		\caption{AFDM fractional delay--Doppler matching-score map.}
		\label{fig:afdm_frac_dd_ridge}
		\vspace{-0.3em}
	\end{figure}
	
	Fig.~\ref{fig:afdm_frac_dd_ridge} illustrates the local ridge
	structure of the fractional-AFDM matching-score surface. The dashed line indicates the
	coupling direction $\Delta\nu=\rho\Delta\ell$, along which
	$\Delta\eta\approx0$, and the elongated high-score region is
	approximately aligned with this direction rather than either
	physical-coordinate axis.
	
	The goal is to estimate $\{\theta_p,\ell_p,\nu_p,\alpha_p\}_{p=1}^{P}$ from $\mathbf{y}$. The angle, delay, and Doppler search bounds, transmitted symbols, and receive-array calibration are assumed to be known. 
	
	\section{Proposed Estimation Framework}
	\label{sec:algorithm}
	
	\subsection{Formulation and Coupled-Coordinate Objective}
	
	\subsubsection{Coarse Sparse Formulation}
	
	Let $\mathcal{G}_{\theta}$, $\mathcal{G}_{\ell}$, and $\mathcal{G}_{\nu}$ denote the angle, delay, and Doppler grids with cardinalities $g_{\theta}$, $g_{\ell}$, and $g_{\nu}$, respectively. Evaluating \eqref{eq:atom_explicit} over all grid points yields the sensing dictionary
	\begin{equation}
		\mathbf{A}\in\mathbb{C}^{N_{\mathrm{R}} N N_{\mathrm c}\times g_{\theta}g_{\ell}g_{\nu}},
	\end{equation}
	whose columns are indexed by the grid triplets $(\theta_o,\ell_q,\nu_r)$, with $\theta_o\in\mathcal{G}_{\theta}$, $\ell_q\in\mathcal{G}_{\ell}$, $\nu_r\in\mathcal{G}_{\nu}$, and column index $i(o,q,r)$. The corresponding column is
	\begin{equation}
		\relax[\mathbf{A}]_{:,i(o,q,r)}
		=\mathbf{a}(\theta_o,\ell_q,\nu_r).
		\label{eq:dictionary_column}
	\end{equation}
	
	For the ideal on-grid case, the sparse coefficient vector $\boldsymbol{\beta}\in\mathbb{C}^{g_{\theta}g_{\ell}g_{\nu}\times 1}$ has nonzero entries only at the true target grid locations. In the considered off-grid setting, this grid-based model is used only for coarse support initialization. The selected supports are then refined in the continuous domain, so the final estimates are not constrained to $\mathcal{G}_{\theta}$, $\mathcal{G}_{\ell}$, and $\mathcal{G}_{\nu}$. The resulting sparse sensing model is
	\begin{equation}
		\mathbf{y}\approx\mathbf{A}\boldsymbol{\beta}+\mathbf{w}.
		\label{eq:grid_sparse_model}
	\end{equation}
	
	For a dense grid, the number of unknown coefficients generally exceeds the number of observations, i.e., $N_{\mathrm{R}} N N_{\mathrm c}\ll g_{\theta}g_{\ell}g_{\nu}$. Hence, \eqref{eq:grid_sparse_model} is underdetermined and can be formulated as
	\begin{equation}
		\min_{\boldsymbol{\beta}}\lVert\boldsymbol{\beta}\rVert_0
		\quad \mathrm{s.t.}\quad
		\lVert\mathbf{y}-\mathbf{A}\boldsymbol{\beta}\rVert_2\leq \varepsilon,
		\label{eq:l0_problem}
	\end{equation}
	where $\varepsilon$ depends on the noise level. The sparse formulation in \eqref{eq:l0_problem} is not solved exactly; it motivates the greedy coarse-support initialization used in Algorithm~\ref{alg:proposed}.
	
	\subsubsection{Off-Grid Refinement Objective}
	Since the target parameters are continuous-valued, the coarse-grid supports are refined in the continuous physical domain. Define the parameter vector of the $p$th target as
	\begin{equation}
		\mathbf{u}_p=[\theta_p,\ell_p,\nu_p]^T .
		\label{eq:u_param}
	\end{equation}
	Given the current estimates, the residual associated with the $p$th target is formed by canceling the contributions of the other initialized targets:
	\begin{equation}
		\mathbf{r}_p
		=
		\mathbf{y}
		-
		\sum_{\substack{i\in\mathcal{S}\\ i\neq p}}
		\widehat{\alpha}_i
		\mathbf{a}(\widehat{\mathbf{u}}_i),
		\label{eq:single_residual}
	\end{equation}
	where $\mathcal{S}$ denotes the current set of initialized targets.
	The single-target refinement is formulated as
	\begin{equation}
		\min_{\alpha_p,\mathbf{u}_p}
		\left\|
		\mathbf{r}_p-\alpha_p\mathbf{a}(\mathbf{u}_p)
		\right\|_2^2 .
		\label{eq:single_target_ls}
	\end{equation}
	For fixed $\mathbf{u}_p$, the least-squares complex coefficient is obtained in closed form as
	\begin{equation}
		\widehat{\alpha}_p(\mathbf{u}_p)
		=
		\frac{\mathbf{a}^H(\mathbf{u}_p)\mathbf{r}_p}
		{\mathbf{a}^H(\mathbf{u}_p)\mathbf{a}(\mathbf{u}_p)} .
		\label{eq:alpha_ls}
	\end{equation}
	Substituting \eqref{eq:alpha_ls} into \eqref{eq:single_target_ls} yields the concentrated objective
	\begin{equation}
		J_{\mathrm{LS}}(\mathbf{u}_p)
		=
		\|\mathbf{r}_p\|_2^2
		-
		\frac{
			|\mathbf{a}^H(\mathbf{u}_p)\mathbf{r}_p|^2
		}
		{
			\|\mathbf{a}(\mathbf{u}_p)\|_2^2
		}.
		\label{eq:profiled_objective}
	\end{equation}
	Since the first term in \eqref{eq:profiled_objective} is independent of $\mathbf{u}_p$, minimizing $J_{\mathrm{LS}}(\mathbf{u}_p)$ is equivalent to maximizing the normalized matching score
	\begin{equation}
		S(\mathbf{u}_p)
		=
		\frac{
			|\mathbf{a}^H(\mathbf{u}_p)\mathbf{r}_p|^2
		}
		{
			\|\mathbf{a}(\mathbf{u}_p)\|_2^2
		}.
		\label{eq:score_u}
	\end{equation}
	
	\subsubsection{Coupled-Coordinate Reparameterization}
	\label{subsubsec:coupled_coordinate_reparam}
	The coupling phase in \eqref{eq:coupling_phase} shows that delay and Doppler deviations affect the DAFT-domain displacement jointly. Using the coupling coefficient $\rho$ defined in Section~II, the coupled coordinate is introduced as
	\begin{equation}
		\mathbf{z}=[\theta,\ell,\eta]^T,\qquad
		\nu=\eta+\rho\,\ell .
		\label{eq:z_param}
	\end{equation}
	Accordingly, define the mapping from the coupled coordinate to the physical coordinate as
	\begin{equation}
		\mathbf{u}(\mathbf{z})
		=
		[\theta,\ell,\eta+\rho\,\ell]^T .
		\label{eq:z_to_u}
	\end{equation}
	The coupled-coordinate score is then given by
	\begin{equation}
		S_z(\mathbf{z})
		\triangleq
		S\big(\mathbf{u}(\mathbf{z})\big).
		\label{eq:score_z}
	\end{equation}

The mapping in \eqref{eq:z_to_u} is an invertible linear
reparameterization and therefore does not by itself alter an
exact unconstrained Newton step. Its role is to expose the
dominant AFDM-induced ridge and provide a coordinate aligned
with this ridge for safeguarded updating, Doppler reconstruction,
and the subsequent coupling-aligned delay refinement.

The dominant DAFT-bin displacement is controlled primarily
by $\eta$, while the residual delay-dependent phases and
inter-symbol Doppler rotation retain separate information
about $\ell$ and $\nu$. The final estimates therefore remain
the physical parameter triplet $(\theta,\ell,\nu)$.
	
	\subsection{Coupled-Coordinate NOMP Estimator}
	\label{sec:method}

	\subsubsection{Orthogonal Matching Pursuit and Local Fine-Grid Initialization}
	
	The orthogonal matching pursuit (OMP) stage provides coarse support
	estimates before the NOMP-type continuous refinement \cite{Tropp2007TIT,Ramasamy2016TSP}. Initialize
	$\mathbf{r}^{(0)}=\mathbf{y}$, $\Lambda_0=\emptyset$, and $\boldsymbol{\Pi}_0=\emptyset$. At the $p$th iteration, the residual is correlated with the unselected dictionary atoms, and the atom index is chosen as
	\begin{equation}
		i_p
		=
		\arg\max_{j\notin\Lambda_{p-1}}
		\frac{
			|\mathbf{A}_j^H\mathbf{r}^{(p-1)}|^2
		}
		{
			\|\mathbf{A}_j\|_2^2
		},
		\label{eq:omp_selection}
	\end{equation}
	where $\mathbf{A}_j$ denotes the $j$th column of $\mathbf{A}$. For a coarse-grid OMP update, the support set and selected-atom matrix are
	\begin{equation}
		\Lambda_p
		=
		\Lambda_{p-1}\cup\{i_p\},
		\qquad
		\boldsymbol{\Pi}_p
		=
		[\boldsymbol{\Pi}_{p-1},\mathbf{A}_{i_p}] .
		\label{eq:omp_support_update}
	\end{equation}
	Given a selected atom matrix, the coefficient vector is estimated by least squares as
	\begin{equation}
		\widehat{\boldsymbol{\alpha}}^{(p)}
		=
		\arg\min_{\boldsymbol{\alpha}}
		\|\mathbf{y}-\boldsymbol{\Pi}_p\boldsymbol{\alpha}\|_2^2
		=
		\boldsymbol{\Pi}_p^{\dagger}\mathbf{y},
		\label{eq:omp_ls}
	\end{equation}
	where $(\cdot)^{\dagger}$ denotes the Moore--Penrose pseudoinverse. The corresponding residual is then updated as
	\begin{equation}
		\mathbf{r}^{(p)}
		=
		\mathbf{y}
		-
		\boldsymbol{\Pi}_p\widehat{\boldsymbol{\alpha}}^{(p)} .
		\label{eq:omp_residual_update}
	\end{equation}
	
	Let $\widehat{\mathbf{u}}_p$ be the coarse parameter estimate produced by OMP. To reduce grid mismatch, a local fine-grid set is formed as
	\begin{equation}
		\mathcal{U}_{p}^{\mathrm{loc}}
		=
		\left\{\widehat{\mathbf{u}}_p+
		[\Delta\theta,\Delta\ell,\Delta\nu]^T\right\},
		\label{eq:local_grid}
	\end{equation}
	where the offsets are taken from centered angle, delay, and Doppler neighborhoods around the selected coarse-grid point. The local fine-grid search uses the normalized correlation criterion in \eqref{eq:omp_selection} over the atoms generated from $\mathcal{U}_{p}^{\mathrm{loc}}$. The local fine-grid maximizer initializes the continuous refinement stage. Once the current support has been refined, the selected dictionary is rebuilt with the refined atoms, and the coefficients are jointly updated by least squares.

	Following coarse OMP detection, subsequent parameter refinements are restricted to the initialized supports, without additional global support search.

	\subsubsection{Coupled-Coordinate Newton Refinement}
	
	Newton refinement is performed in the coupled coordinate introduced in Section~\ref{subsubsec:coupled_coordinate_reparam}. Given the current physical-domain estimate
	$\widehat{\mathbf{u}}_p=[\widehat{\theta}_p,\widehat{\ell}_p,\widehat{\nu}_p]^T$,
	its coupled-coordinate representation is
	\begin{equation}
		\widehat{\mathbf{z}}_p
		=
		[
		\widehat{\theta}_p,
		\widehat{\ell}_p,
		\widehat{\eta}_p
		]^T,
		\qquad
		\widehat{\eta}_p
		=
		\widehat{\nu}_p-\rho\widehat{\ell}_p .
		\label{eq:u_to_z_newton}
	\end{equation}
	Starting from $\mathbf{z}^{(0)}=\widehat{\mathbf{z}}_p$, the score
	$S_z(\mathbf{z})$ is locally approximated at the $t$th iteration by
	\begin{equation}
		S_z(\mathbf{z}^{(t)}+\Delta\mathbf{z})
		\approx
		S_z(\mathbf{z}^{(t)})
		+
		\mathbf{g}_t^T\Delta\mathbf{z}
		+
		\frac{1}{2}
		\Delta\mathbf{z}^T
		\mathbf{H}_t
		\Delta\mathbf{z},
		\label{eq:newton_quadratic}
	\end{equation}
	where
	\begin{equation}
		\mathbf{g}_t
		=
		\nabla_{\mathbf{z}}S_z(\mathbf{z}^{(t)}),
		\qquad
		\mathbf{H}_t
		=
		\nabla_{\mathbf{z}}^2S_z(\mathbf{z}^{(t)}).
		\label{eq:gradient_hessian}
	\end{equation}
	The stationary condition of the local quadratic model gives
	\begin{equation}
		\mathbf{g}_t+\mathbf{H}_t\Delta\mathbf{z}_t=\mathbf{0},
		\label{eq:newton_stationary}
	\end{equation}
	and the Newton direction is computed as
	\begin{equation}
		\Delta\mathbf{z}_t
		=
		-\mathbf{H}_t^{\dagger}\mathbf{g}_t,
		\label{eq:newton_step}
	\end{equation}
	where \((\cdot)^\dagger\) is used for ill-conditioned local quadratic
	models. Feasibility and monotonicity are enforced by the safeguarded
	update below.
	
	A safeguarded update is then applied as
	\begin{equation}
		\mathbf{z}^{(t+1)}
		=
		\mathcal{P}_{\Omega_z}
		\left(
		\mathbf{z}^{(t)}
		+
		\mu_t\Delta\mathbf{z}_t
		\right),
		\label{eq:safeguarded_newton}
	\end{equation}
	where $\mu_t\in(0,1]$ is selected by backtracking and
	$\mathcal{P}_{\Omega_z}(\cdot)$ denotes a boundary feasibility mapping associated with the feasible set
	\begin{equation}
		\begin{aligned}
			\Omega_z
			=
			\{(\theta,\ell,\eta):\;&
			\theta_{\mathrm{min}}\leq\theta\leq\theta_{\mathrm{max}},\;
			\ell_{\mathrm{min}}\leq\ell\leq\ell_{\mathrm{max}},\\
			&
			\nu_{\mathrm{min}}
			\leq
			\eta+\rho\,\ell
			\leq
			\nu_{\mathrm{max}}
			\}.
		\end{aligned}
		\label{eq:feasible_z_set}
	\end{equation}
	The correction enforces the feasible physical bounds through the mapping $\nu=\eta+\rho\ell$. Specifically, the coupled coordinate $\mathbf{z}$ is first converted to the physical coordinate $\mathbf{u}$, then clipped to the feasible angle--delay--Doppler region, and finally mapped back to the coupled coordinate. The candidate update is accepted only if it does not decrease $S_z(\mathbf{z})$; otherwise, the step size $\mu_t$ is reduced by backtracking.

	The score gradient and Hessian in \eqref{eq:gradient_hessian} are evaluated using the analytical expressions summarized in Appendix~\ref{app:atom_derivatives}. These expressions use the first- and second-order derivatives of the AFDM atom and the coupled-coordinate chain rule for $\mathbf{z}=[\theta,\ell,\eta]^T$.
	
	Let $\widehat{\mathbf{z}}_p$ denote the accepted coupled-coordinate estimate after Newton refinement. It is converted back to the physical domain as
	\begin{equation}
		\widehat{\theta}_p=\widehat{z}_{p,1},
		\qquad
		\widehat{\ell}_p=\widehat{z}_{p,2},
		\qquad
		\widehat{\nu}_p=\widehat{z}_{p,3}+\rho\widehat{z}_{p,2}.
		\label{eq:z_to_u_after_newton}
	\end{equation}
	The converted estimate is then further refined by a coordinate-wise local search over $(\theta,\ell,\nu)$ using the score in \eqref{eq:score_u}. Among the coarse initial point, the coupled-coordinate
	refined point, and the physical-coordinate refined point, the
	candidate with the largest score is retained for
	coupling-aligned delay refinement.
	
	\subsubsection{Coupling-Aligned Delay Refinement}
	
	After coupled-coordinate Newton refinement, the proposed coupling-aligned delay refinement fixes $\widehat{\theta}_p$ and
	$\widehat{\eta}_p$, and searches only along the delay coordinate. According to
	the coupling relation $\nu=\eta+\rho\,\ell$, each candidate delay $\ell'$ uniquely
	determines a normalized Doppler candidate
	\begin{equation}
		\nu'=\widehat{\eta}_p+\rho\,\ell' .
		\label{eq:ridge_doppler_candidate}
	\end{equation}
	
	The candidate delay set is centered at the current estimate:
	\[
	\mathcal{L}_{\mathrm{ridge}}
	=\left\{\ell_q^{(\mathrm{r})}\right\}_{q=1}^{G_{\ell,\mathrm{ridge}}}
	\subset
	\left[\widehat{\ell}_p-\Delta\ell_{\mathrm{ridge}},
	\widehat{\ell}_p+\Delta\ell_{\mathrm{ridge}}\right],
	\]
	where $\Delta\ell_{\mathrm{ridge}}$ is the half-width of the local ridge search and $G_{\ell,\mathrm{ridge}}$ is the nominal number of delay candidates before feasibility screening. Candidate points that would place the reconstructed normalized Doppler outside the feasible interval are removed. The selected delay maximizes the residual score along the ridge:
	\begin{equation}
		\widehat{\ell}_p
		=
		\operatorname*{arg\,max}_{\ell'\in\mathcal{L}_{\mathrm{ridge}}}
		S\!\left(\widehat{\theta}_p,\ell',
		\widehat{\eta}_p+\rho\,\ell'\right).
		\label{eq:ridge_search}
	\end{equation}
	The normalized Doppler estimate is then updated as
	\begin{equation}
		\widehat{\nu}_p=\widehat{\eta}_p+\rho\widehat{\ell}_p.
	\end{equation}
	Thus, coupling-aligned delay refinement performs a one-dimensional local correction along the AFDM-induced coupling direction, avoiding a full three-dimensional local search.

	\subsubsection{Cyclic Multi-Target Refinement}
	Once all $P$ targets have been initialized, cyclic refinement is performed to reduce inter-target leakage and error propagation. For the $p$th target, an interference-reduced residual is constructed by removing the contributions of the other currently estimated targets:
	\begin{equation}
		\mathbf{r}'_p
		=
		\mathbf{y}
		-
		\sum_{i=1,\,i\neq p}^{P}
		\widehat{\alpha}_i
		\mathbf{a}(\widehat{\mathbf{u}}_i).
		\label{eq:cyclic_residual}
	\end{equation}
	Using $\mathbf{r}'_p$, coupled-coordinate Newton refinement, local physical-coordinate refinement, and coupling-aligned delay refinement are reapplied to the $p$th target. The refined dictionary is then rebuilt, and the reflection coefficients are jointly updated by least squares.
	
	One cyclic refinement round processes all $P$ targets once. The simulations use a fixed number of cyclic refinement rounds, although a residual-reduction stopping rule can also be used.
	
	Upon completion of cyclic refinement, all reflection coefficients are jointly updated as
	\begin{equation}
		\widehat{\boldsymbol{\alpha}}
		=
		\arg\min_{\boldsymbol{\alpha}}
		\left\|
		\mathbf{y}
		-
		\widehat{\mathbf{A}}\boldsymbol{\alpha}
		\right\|_2^2.
		\label{eq:joint_ls}
	\end{equation}
	The final output is
	$\{\widehat{\theta}_p,\widehat{\ell}_p,\widehat{\nu}_p,\widehat{\alpha}_p\}_{p=1}^{P}$.
	
Relative to Standard 3D NOMP, which directly refines $(\theta,\ell,\nu)$, CC-NOMP introduces the coupled coordinate $(\theta,\ell,\eta)$ with $\eta=\nu-\rho\ell$ and then applies coupling-aligned delay refinement.

\begin{algorithm}[!t]
	\caption{Coupled-Coordinate NOMP Estimator (CC-NOMP)}
	\label{alg:proposed}
	\begin{algorithmic}[1]
		\REQUIRE Observation $\mathbf{y}$, coarse dictionary $\mathbf{A}$, target number $P$, cyclic refinement rounds $I_{\mathrm{C}}$, optional threshold $\epsilon_{\mathrm{cyc}}$
		\ENSURE $\{\widehat{\theta}_p,\widehat{\ell}_p,\widehat{\nu}_p,\widehat{\alpha}_p\}_{p=1}^{P}$
		
		\STATE Initialize $\mathbf{r}^{(0)}=\mathbf{y}$, $\widehat{\mathbf{A}}_0=\emptyset$, and $\Lambda_0=\emptyset$.
		
		\FOR{$p=1,\cdots,P$}
		\STATE Select the coarse index $i_p$ by \eqref{eq:omp_selection} over $j\notin\Lambda_{p-1}$ using $\mathbf{r}^{(p-1)}$.
		\STATE Obtain the coarse estimate $\widehat{\mathbf{u}}_p^{\mathrm{c}}$ associated with $i_p$.
		\STATE Update $\Lambda_p=\Lambda_{p-1}\cup\{i_p\}$.
		\STATE Obtain $\widehat{\mathbf{u}}_p^{(0)}$ by the local fine-grid search in \eqref{eq:local_grid}.
		\STATE Convert $\widehat{\mathbf{u}}_p^{(0)}$ to the coupled coordinate $\widehat{\mathbf{z}}_p^{(0)}$.
		\STATE Apply coupled-coordinate Newton refinement by \eqref{eq:newton_step} and \eqref{eq:safeguarded_newton}.
		\STATE Apply local physical-coordinate refinement and retain the candidate with the largest normalized matching score.
		\STATE Apply coupling-aligned delay refinement by \eqref{eq:ridge_search} and recover $\widehat{\nu}_p=\widehat{\eta}_p+\rho\widehat{\ell}_p$.
		\STATE Add the refined atom to $\widehat{\mathbf{A}}_p=[\widehat{\mathbf{A}}_{p-1},\mathbf{a}(\widehat{\mathbf{u}}_p)]$.
		\STATE Jointly update $\{\widehat{\alpha}_i\}_{i=1}^{p}$ by least squares using $\widehat{\mathbf{A}}_p$.
		\STATE Refresh $\mathbf{r}^{(p)}=\mathbf{y}-\widehat{\mathbf{A}}_p\widehat{\boldsymbol{\alpha}}^{(p)}$ before detecting the next target.
		\ENDFOR
		
		\FOR{$i_{\mathrm{c}}=1,\cdots,I_{\mathrm{C}}$}
		\STATE Set $\mathbf{r}_{\mathrm{old}}=\mathbf{y}-\widehat{\mathbf{A}}\widehat{\boldsymbol{\alpha}}$.
		\FOR{$p=1,\cdots,P$}
		\STATE Construct $\mathbf{r}'_p$ by \eqref{eq:cyclic_residual}.
		\STATE Reapply Lines 7--10 using $\mathbf r'_p$, initialized from
		the current estimate $\hat{\mathbf u}_p$.
		\STATE Rebuild the current refined dictionary and update the reflection coefficients by least squares.
		\ENDFOR
		\STATE Set $\mathbf{r}_{\mathrm{new}}=\mathbf{y}-\widehat{\mathbf{A}}\widehat{\boldsymbol{\alpha}}$ and compute the residual-energy decrease $\Delta r=\|r_{\rm old}\|_2^2-\|r_{\rm new}\|_2^2$.
		\IF{$\epsilon_{\mathrm{cyc}}>0$ and $\Delta r/\lVert\mathbf{r}_{\mathrm{old}}\rVert_2^2<\epsilon_{\mathrm{cyc}}$}
		\STATE \textbf{break}
		\ENDIF
		\ENDFOR
		
		\STATE Jointly update $\{\hat{\alpha}_p\}_{p=1}^{P}$ by the final least-squares update.
		\RETURN $\{\widehat{\theta}_p,\widehat{\ell}_p,\widehat{\nu}_p,\widehat{\alpha}_p\}_{p=1}^{P}$.
	\end{algorithmic}
\end{algorithm}

The proposed CC-NOMP estimator refines coarse OMP supports through
local initialization, coupled-coordinate Newton refinement,
coupling-aligned delay refinement, and cyclic multi-target refinement.
The auxiliary coordinate parameterizes the dominant ridge in the
AFDM delay--Doppler matching-score surface, while coupling-aligned
delay refinement performs a one-dimensional correction along this ridge.
Algorithm~\ref{alg:proposed} summarizes the proposed CC-NOMP
estimator.

	\section{Performance Analysis}
	\label{sec:performance_analysis}
	
	This section first derives a deterministic CRB benchmark for the continuous sensing parameters and then analyzes the dominant computational complexity of the proposed CC-NOMP estimator and the considered baselines.

	\subsection{CRB Derivation}
	\label{subsec:crb}
	
	The deterministic CRB is derived under the same DAFT-domain fractional response model and known-symbol assumption used by the estimator. It should therefore be interpreted as an ideal matched-model local lower bound. The stacked sensing model in \eqref{eq:stacked_model} can be equivalently written as
	\begin{equation}
		\mathbf{y}
		=
		\sum_{p=1}^{P}
		\alpha_p
		\mathbf{a}(\theta_p,\ell_p,\nu_p)
		+
		\mathbf{w}
		=
		\mathbf{A}(\boldsymbol{\chi})\boldsymbol{\alpha}
		+
		\mathbf{w},
		\label{eq:crb_stacked_model}
	\end{equation}
	where the unknown sensing-parameter vector is defined as
	\begin{equation}
		\boldsymbol{\chi}
		=
		[
		\theta_1,\ell_1,\nu_1,\cdots,\theta_P,\ell_P,\nu_P
		]^T .
		\label{eq:crb_chi}
	\end{equation}
	The complex reflection coefficients are treated as deterministic nuisance parameters. The corresponding real-valued parameter vector is
	\begin{equation}
		\boldsymbol{\vartheta}
		=
		[
		\boldsymbol{\chi}^T,
		\operatorname{Re}\{\boldsymbol{\alpha}\}^T,
		\operatorname{Im}\{\boldsymbol{\alpha}\}^T
		]^T
		\in\mathbb{R}^{5P}.
		\label{eq:crb_vartheta}
	\end{equation}
	The noise follows
	\begin{equation}
		\mathbf{w}\sim
		\mathcal{CN}
		(
		\mathbf{0},
		\sigma_w^2\mathbf{I}_{N_{\mathrm{R}}NN_{\mathrm c}}
		).
		\label{eq:crb_noise}
	\end{equation}
	If $\mathbf{y}_0=\mathbf{A}(\boldsymbol{\chi})\boldsymbol{\alpha}$ denotes the noiseless DAFT-domain observation and $M_{\mathrm{obs}}=N_{\mathrm{R}}NN_{\mathrm c}$, then
	\begin{equation}
		\mathrm{SNR}
		=
		10\log_{10}
		\left(
		\frac{\lVert\mathbf{y}_0\rVert_2^2/M_{\mathrm{obs}}}{\sigma_w^2}
		\right).
		\label{eq:snr_definition}
	\end{equation}
	This signal-to-noise ratio (SNR) is defined per stacked DAFT-domain observation sample after the adopted DAFT-domain normalization. Since the DAFT operation is unitary, the average noise power is preserved between the useful time-domain samples and the DAFT-domain samples. The same SNR definition is used for Monte Carlo noise generation and deterministic CRB evaluation. For a complex Gaussian observation with parameter-independent covariance, the Fisher information matrix (FIM) is given by
	\begin{equation}
		[
		\mathbf{J}(\boldsymbol{\vartheta})
		]_{i,j}
		=
		\frac{2}{\sigma_w^2}
		\operatorname{Re}
		\left\{
		\left(
		\frac{\partial\boldsymbol{\mu}}
		{\partial\vartheta_i}
		\right)^H
		\left(
		\frac{\partial\boldsymbol{\mu}}
		{\partial\vartheta_j}
		\right)
		\right\},
		\label{eq:crb_fim_entry}
	\end{equation}
	where
	\begin{equation}
		\boldsymbol{\mu}(\boldsymbol{\vartheta})
		=
		\mathbf{A}(\boldsymbol{\chi})\boldsymbol{\alpha}.
		\label{eq:crb_mean}
	\end{equation}
	Let $\mathbf{a}_p=\mathbf{a}(\theta_p,\ell_p,\nu_p)$. The required mean-vector derivatives are
	\begin{equation}
		\begin{aligned}
			&\frac{\partial\boldsymbol{\mu}}{\partial\theta_p}
			=
			\alpha_p\frac{\partial\mathbf{a}_p}{\partial\theta_p},
			&
			&\frac{\partial\boldsymbol{\mu}}{\partial\ell_p}
			=
			\alpha_p\frac{\partial\mathbf{a}_p}{\partial\ell_p},
			\\
			&\frac{\partial\boldsymbol{\mu}}{\partial\nu_p}
			=
			\alpha_p\frac{\partial\mathbf{a}_p}{\partial\nu_p},
			&
			&\frac{\partial\boldsymbol{\mu}}
			{\partial\operatorname{Re}\{\alpha_p\}}
			=
			\mathbf{a}_p,
			\\
			&\frac{\partial\boldsymbol{\mu}}
			{\partial\operatorname{Im}\{\alpha_p\}}
			=
			\jj\mathbf{a}_p .
		\end{aligned}
		\label{eq:crb_derivatives}
	\end{equation}
	The atom derivatives with respect to $\theta_p$, $\ell_p$, and $\nu_p$ are evaluated from the analytical expressions in Appendix~\ref{app:atom_derivatives}.
	
	Define the derivative matrix as
	\begin{equation}
		\mathbf{D}_{\boldsymbol{\vartheta}}
		=
		\left[
		\frac{\partial\boldsymbol{\mu}}{\partial\vartheta_1},
		\frac{\partial\boldsymbol{\mu}}{\partial\vartheta_2},
		\cdots,
		\frac{\partial\boldsymbol{\mu}}{\partial\vartheta_{5P}}
		\right].
		\label{eq:crb_derivative_matrix}
	\end{equation}
	Then, the compact FIM expression is
	\begin{equation}
		\mathbf{J}(\boldsymbol{\vartheta})
		=
		\frac{2}{\sigma_w^2}
		\operatorname{Re}
		\left\{
		\mathbf{D}_{\boldsymbol{\vartheta}}^H\mathbf{D}_{\boldsymbol{\vartheta}}
		\right\}.
		\label{eq:crb_fim_compact}
	\end{equation}
	Partition the FIM according to the parameters of interest and nuisance parameters:
	\begin{equation}
		\mathbf{J}
		=
		\begin{bmatrix}
			\mathbf{J}_{\boldsymbol{\chi}\boldsymbol{\chi}} & \mathbf{J}_{\boldsymbol{\chi}\boldsymbol{\alpha}} \\
			\mathbf{J}_{\boldsymbol{\alpha}\boldsymbol{\chi}} & \mathbf{J}_{\boldsymbol{\alpha}\boldsymbol{\alpha}}
		\end{bmatrix}.
		\label{eq:crb_fim_partition}
	\end{equation}
	The equivalent CRB for $\boldsymbol{\chi}$ is obtained from the Schur complement as
	\begin{equation}
		\mathbf{C}_{\boldsymbol{\chi}}
		=
		\left(
		\mathbf{J}_{\boldsymbol{\chi}\boldsymbol{\chi}}
		-
		\mathbf{J}_{\boldsymbol{\chi}\boldsymbol{\alpha}}
		\mathbf{J}_{\boldsymbol{\alpha}\boldsymbol{\alpha}}^{-1}
		\mathbf{J}_{\boldsymbol{\alpha}\boldsymbol{\chi}}
		\right)^{-1}.
		\label{eq:crb_equivalent}
	\end{equation}
	When $\mathbf{J}_{\boldsymbol{\alpha}\boldsymbol{\alpha}}$ or the Schur complement is ill-conditioned, the Moore--Penrose inverse is used with the same numerical inversion rule for all SNR points and Monte Carlo target realizations.
	
	Let $\mathcal{I}_{\theta}$, $\mathcal{I}_{\ell}$, and $\mathcal{I}_{\nu}$ denote the index sets of angle, normalized delay, and normalized Doppler entries in $\boldsymbol{\chi}$, respectively. The corresponding average CRBs are
	\begin{equation}
		\begin{aligned}
			\mathrm{CRB}_{\theta}
			&=
			\frac{1}{P}
			\operatorname{Tr}
			\left(
			[
			\mathbf{C}_{\boldsymbol{\chi}}
			]_{\mathcal{I}_{\theta},\mathcal{I}_{\theta}}
			\right),
			\\
			\mathrm{CRB}_{\ell}
			&=
			\frac{1}{P}
			\operatorname{Tr}
			\left(
			[
			\mathbf{C}_{\boldsymbol{\chi}}
			]_{\mathcal{I}_{\ell},\mathcal{I}_{\ell}}
			\right),
			\\
			\mathrm{CRB}_{\nu}
			&=
			\frac{1}{P}
			\operatorname{Tr}
			\left(
			[
			\mathbf{C}_{\boldsymbol{\chi}}
			]_{\mathcal{I}_{\nu},\mathcal{I}_{\nu}}
			\right).
		\end{aligned}
		\label{eq:crb_average}
	\end{equation}
	The root-mean-square error (RMSE) lower bounds are therefore
	\begin{equation}
		\begin{aligned}
			\mathrm{RMSE}_{\theta,\mathrm{CRB}}
			&=
			\frac{180}{\pi}
			\sqrt{\mathrm{CRB}_{\theta}},
			\\
			\mathrm{RMSE}_{\ell,\mathrm{CRB}}
			&=
			\sqrt{\mathrm{CRB}_{\ell}},
			\\
			\mathrm{RMSE}_{\nu,\mathrm{CRB}}
			&=
			\sqrt{\mathrm{CRB}_{\nu}}.
		\end{aligned}
		\label{eq:crb_rmse_bounds}
	\end{equation}
	The factor $180/\pi$ in \eqref{eq:crb_rmse_bounds} converts the angular RMSE lower bound from radians to degrees, consistent with the angle RMSE curves in Section~V.
	
	The CRB of the coupled coordinate can be obtained by error propagation. From
	\begin{equation}
		\eta_p=\nu_p-\rho\,\ell_p,
		\qquad
		\rho=2N_{\mathrm c}\,c_1,
		\label{eq:crb_eta_def}
	\end{equation}
	the perturbation relation is
	\begin{equation}
		\Delta\eta_p
		=
		\Delta\nu_p
		-
		\rho\Delta\ell_p .
		\label{eq:crb_eta_error}
	\end{equation}
	Define
	\begin{equation}
		i_{\ell,p}=3p-1,\qquad i_{\nu,p}=3p,
		\label{eq:crb_eta_indices}
	\end{equation}
	where $i_{\ell,p}$ and $i_{\nu,p}$ denote the positions of $\ell_p$ and $\nu_p$ in $\boldsymbol{\chi}$, respectively.
	Thus,
	\begin{equation}
		\begin{aligned}
			\mathrm{CRB}_{\eta_p}
			&=
			[
			\mathbf{C}_{\boldsymbol{\chi}}
			]_{i_{\nu,p},i_{\nu,p}}
			+
			\rho^2
			[
			\mathbf{C}_{\boldsymbol{\chi}}
			]_{i_{\ell,p},i_{\ell,p}}
			\\
			&\quad -
			2\rho
			[
			\mathbf{C}_{\boldsymbol{\chi}}
			]_{i_{\ell,p},i_{\nu,p}}.
		\end{aligned}
		\label{eq:crb_eta_single}
	\end{equation}
	The average coupled-coordinate CRB is
	\begin{equation}
		\mathrm{CRB}_{\eta}
		=
		\frac{1}{P}
		\sum_{p=1}^{P}
		\mathrm{CRB}_{\eta_p},
		\label{eq:crb_eta_average}
	\end{equation}
	with the corresponding RMSE lower bound
	\begin{equation}
		\mathrm{RMSE}_{\eta,\mathrm{CRB}}
		=
		\sqrt{\mathrm{CRB}_{\eta}}.
		\label{eq:crb_eta_rmse}
	\end{equation}
	The numerical CRB is evaluated from the stacked
	AFDM atom model at the true target parameters and
	reflection coefficients of each Monte Carlo target realization, using
	the same transmitted QAM symbols and SNR definition as in
	the Monte Carlo experiments. The reflection coefficients are treated as nuisance parameters and eliminated by the Schur complement, with a Moore--Penrose inverse used only for numerical stabilization when needed. For each SNR point, the CRB variances are averaged over the target realizations and labeled targets before taking the square root. Because the empirical RMSE uses permutation matching whereas
	the CRB is defined for locally labeled parameters, the CRB
	comparison is interpreted primarily in the medium- and
	high-SNR local-estimation regime.
	
	\subsection{Complexity Analysis}
	\label{subsec:complexity}
	
	Let $M_{\mathrm{obs}}=N_{\mathrm{R}}NN_{\mathrm c}$ denote the stacked observation length, and let $G_{\mathrm{c}}=g_{\theta}g_{\ell}g_{\nu}$ denote the coarse dictionary size. Here, $P$, $I_{\mathrm{C}}$, and $I_{\mathrm{N}}$ denote the target number, the number of cyclic refinement rounds, and the number of Newton iterations, respectively. For Fractional Local maximum-likelihood (ML), $I_{\rm F}$ denotes the number of
	cyclic refinement rounds, and the local search grid size is denoted by $G_{\mathrm{loc}}=G_{\theta,\mathrm{loc}}G_{\ell,\mathrm{loc}}G_{\nu,\mathrm{loc}}$. For CC-NOMP, $G_{\ell,\mathrm{ridge}}$ denotes the number of one-dimensional candidates used in coupling-aligned delay refinement. The fixed-size local fine-grid initialization in Standard 3D
	NOMP and CC-NOMP and the local physical-coordinate search
	in CC-NOMP are omitted from the dominant-order table. The factors $1+I_{\mathrm C}$ and $1+I_{\mathrm F}$
	account for one initial refinement and $I_{\mathrm C}$ or
	$I_{\mathrm F}$ subsequent cyclic refinement rounds, respectively.

	For Coarse-grid OMP, the dominant complexity is
	\begin{equation}
		\mathcal{O}
		\left(
		PM_{\mathrm{obs}}G_{\mathrm{c}}
		+M_{\mathrm{obs}}P^2
		+P^3
		\right),
		\label{eq:complexity_omp}
	\end{equation}
	where $PM_{\mathrm{obs}}G_{\mathrm{c}}$ comes from residual correlations with the coarse dictionary. For Standard 3D NOMP, the dominant complexity is
	\begin{equation}
		\mathcal{O}
		\left(
		PM_{\mathrm{obs}}G_{\mathrm{c}}
		+P(1+I_{\mathrm{C}})I_{\mathrm{N}}M_{\mathrm{obs}}
		+M_{\mathrm{obs}}P^2
		+P^3
		\right),
		\label{eq:complexity_nomp}
	\end{equation}
	where the second term accounts for local Newton refinement during the initial refinement and the subsequent cyclic refinement rounds. For Fractional Local ML, the dominant complexity is
	\begin{equation}
		\mathcal{O}
		\left(
		PM_{\mathrm{obs}}G_{\mathrm{c}}
		+P(1+I_{\rm F})M_{\mathrm{obs}}G_{\mathrm{loc}}
		+M_{\mathrm{obs}}P^2
		+P^3
		\right).
		\label{eq:complexity_local_ml}
	\end{equation}
	Its cost depends on the adopted local search window and grid density. For the proposed CC-NOMP estimator, the dominant complexity is
	\begin{equation}
		\begin{aligned}
			\mathcal{O}\Big(
			&PM_{\mathrm{obs}}G_{\mathrm{c}}
			+P(1+I_{\mathrm{C}})I_{\mathrm{N}}M_{\mathrm{obs}}
			\\
			&+P(1+I_{\mathrm{C}})G_{\ell,\mathrm{ridge}}M_{\mathrm{obs}}
			\\
			&+M_{\mathrm{obs}}P^2
			+P^3
			\Big).
		\end{aligned}
	\end{equation}

	For compactness, the least-squares terms \(M_{\mathrm{obs}}P^2+P^3\) are omitted in Table~\ref{tab:complexity}, since \(P\) is small in the considered sensing setup.

All methods share the coarse-search cost
$\mathcal{O}(PM_{\rm obs}G_c)$. Standard 3D NOMP and
CC-NOMP have the same Newton-refinement order. The
additional dominant-order term of CC-NOMP is a
one-dimensional ridge search, whereas Fractional Local ML
performs three-dimensional local searches during both initial
and cyclic refinement.
	\begin{table}[H]
	\caption{Computational Complexity Comparison}
	\label{tab:complexity}
	\centering
	\footnotesize
	\setlength{\tabcolsep}{2pt}
	\begin{tabular}{@{}p{0.30\linewidth}p{0.62\linewidth}@{}}
		\hline
		Method & Dominant complexity \\
		\hline
		Coarse-grid OMP
		&
		\(\mathcal{O}\!\left(PM_{\mathrm{obs}}G_{\mathrm{c}}\right)\)
		\\
		Standard 3D NOMP
		&
		\(\begin{aligned}[t]
			\mathcal{O}\!\big(PM_{\mathrm{obs}}G_{\mathrm{c}}
			+P(1+I_{\mathrm{C}})I_{\mathrm{N}}M_{\mathrm{obs}}\big)
		\end{aligned}\)
		\\
		Fractional Local ML
		&
		\(\mathcal{O}\!\left(PM_{\mathrm{obs}}G_{\mathrm{c}}
		+P(1+I_{\rm F})M_{\mathrm{obs}}G_{\mathrm{loc}}\right)\)
		\\
		CC-NOMP
		&
		\(\begin{aligned}[t]
			\mathcal{O}\!\big(PM_{\mathrm{obs}}G_{\mathrm{c}}
			&+P(1+I_{\mathrm{C}})I_{\mathrm{N}}M_{\mathrm{obs}}\\
			&+P(1+I_{\mathrm{C}})G_{\ell,\mathrm{ridge}}M_{\mathrm{obs}}\big)
		\end{aligned}\)
		\\
		\hline
	\end{tabular}
\end{table}
	\FloatBarrier
	
		\begin{table}[H]
		\caption{Simulation Parameters}
		\label{tab:sim_params}
		\centering
		\footnotesize
		\setlength{\tabcolsep}{2pt}
		\begin{tabular}{@{}p{0.50\columnwidth}p{0.42\columnwidth}@{}}
			\toprule
			Parameter & Value \\
			\midrule
			Carrier frequency & \(f_c=15\) GHz \\
			Bandwidth & \(B=10\) MHz \\
			Transmit configuration & Single transmit antenna \\
			Receive antennas & \(N_{\mathrm{R}}=4\) \\
			Subcarriers & \(N_{\mathrm c}=64\) \\
			AFDM symbols & \(N=4\) \\
			CPP length & \(L_{\mathrm{CPP}}=8\) \\
			Guard width & \(G_{\mathrm{guard}}=2\) \\
			Doppler grid bound & \(v_{\max}=3\) \\
			AFDM chirp parameter \(c_1\) & \((2(v_{\max}+G_{\mathrm{guard}})+1)/(2N_{\mathrm c})\) \\
			AFDM chirp parameter \(c_2\) & \(1/(\pi N_{\mathrm c}^2)\) \\
			Coupling coefficient & \(\rho=2N_{\mathrm c}\,c_1=11\) \\
			Modulation & 16-QAM, unit average power \\
			Reflection coefficients & \(|\alpha_p|\in[0.65,1]\), random phase \\
			Number of targets & \(P=3\) \\
			Angle range & \([-25^\circ,25^\circ]\) \\
			Normalized delay range & \([0.5,7.5]\) \\
			Normalized Doppler range & \([-2.5,2.5]\) \\
			SNR range & $-10$ to $30$ dB with a $2$-dB step \\
			Main SNR-domain trials & 500 Monte Carlo trials \\
			Ablation study trials & 500 Monte Carlo trials at SNR = 10 dB \\
			Target-number sensitivity trials & 500 Monte Carlo trials for each \(P\) \\
			\bottomrule
		\end{tabular}
	\end{table}
	
	\section{NUMERICAL RESULTS}
	\label{sec:simulation}

	\subsection{Simulation Setup and Evaluation Metrics}
	\label{subsec:simulation_setup}
	
	This section evaluates CC-NOMP under the adopted AFDM sensing model.
	All estimators use the same continuous-valued target realization in
	each Monte Carlo trial, with independent noise across SNR points and
	trials. Table~\ref{tab:sim_params} lists the main simulation
	parameters. A candidate target pair is rejected if
	$|\Delta\theta|<1^\circ$, $|\Delta\eta|<0.60$, or
	$(\Delta\theta/2^\circ)^2+(\Delta\ell/0.20)^2+
	(\Delta\nu/0.20)^2<1$. The delay and Doppler parameters are off-grid.
	The received echoes and sensing atoms use the same fractional
	DAFT-domain response, and all estimators share the waveform setting
	and noise convention.
	
	The compared estimators are Coarse-grid OMP, Standard 3D NOMP adapted
	from \cite{Ramasamy2016TSP}, a model-matched Fractional Local ML
	baseline inspired by \cite{Zhu2026WCL}, and CC-NOMP. Standard 3D NOMP
	and CC-NOMP use the same coarse dictionary, parameter bounds,
	coefficient updates, $I_{\mathrm N}=12$ Newton iterations, and
	$I_{\mathrm C}=10$ cyclic refinement rounds. The former refines
	directly in $(\theta,\ell,\nu)$, whereas the latter uses
	coupled-coordinate Newton refinement and coupling-aligned delay
	refinement. Their coarse and local fine grids contain
	$121\times33\times25$ and $13\times11\times11$ points, respectively.
	CC-NOMP uses ridge offsets $\{-0.08,0,0.08\}$, while Fractional Local
	ML performs a $7\times26\times61$ local three-dimensional search
	during the initial refinement and two cyclic refinement rounds with
	joint least-squares coefficient re-estimation. The deterministic CRB
	serves as a matched-model local reference.
	
				\begin{figure*}[!t]
		\centering
		\subfloat[Angle]{%
			\includegraphics[width=0.465\textwidth]{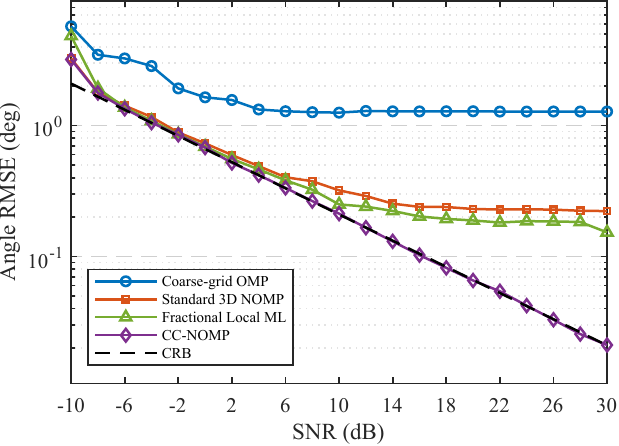}}
		\hfil
		\subfloat[Normalized delay]{%
			\includegraphics[width=0.465\textwidth]{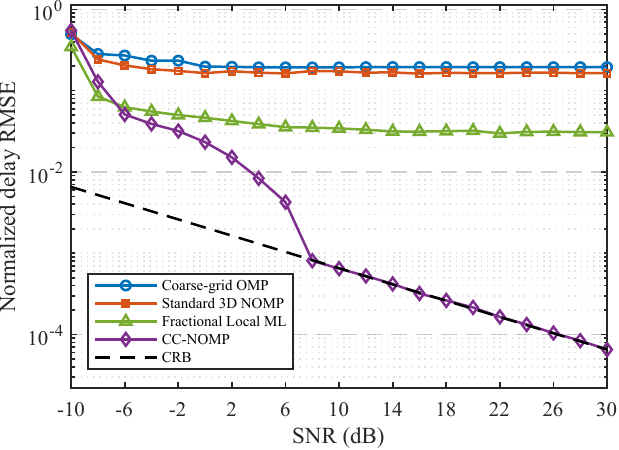}}
		\\[-1mm]
		\subfloat[Normalized Doppler]{%
			\includegraphics[width=0.465\textwidth]{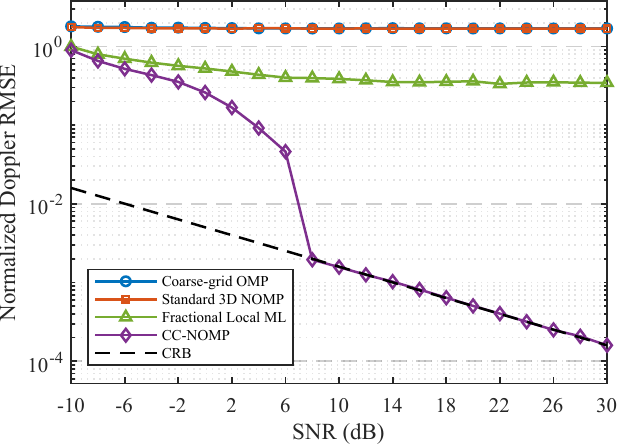}}
		\hfil
		\subfloat[Auxiliary coupling coordinate]{%
			\includegraphics[width=0.465\textwidth]{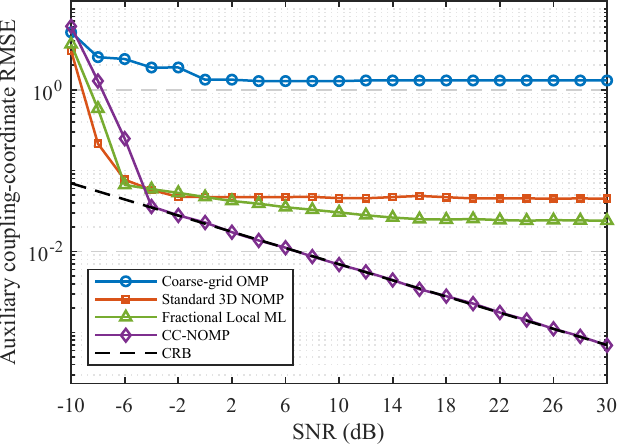}}
		\caption{RMSE performance versus SNR: (a) angle, (b) normalized delay, (c) normalized Doppler, and (d) auxiliary coupling coordinate.}
		\label{fig:main_rmse}
	\end{figure*}
	
	In addition to the physical-parameter RMSEs, the RMSE of
	$\eta=\nu-\rho\ell$ is reported as an auxiliary diagnostic metric.
	Target matching uses an $\eta$-aware permutation distance, with the
	resulting common assignment applied to the angle, delay, Doppler, and
	auxiliary-coordinate errors. For each matched target,
	$\hat{\eta}=\hat{\nu}-\rho\hat{\ell}$. The angle RMSE is reported in
	degrees, while the delay, Doppler, and auxiliary-coordinate RMSEs are
	reported in normalized units.

	\subsection{Estimation Performance and Runtime Comparison}
	\label{subsec:main_snr_runtime}

	Fig.~\ref{fig:main_rmse}(a) shows the angle RMSE versus SNR. Under the adopted receive-array configuration, angular information	is mainly determined by the receive-array spatial phase, and the multi-antenna
	estimators use the same receive-array observation structure for
	angle estimation. As a result, the refined estimators achieve lower
	angle RMSE than Coarse-grid OMP at medium and high SNRs, and
	CC-NOMP maintains angle-estimation accuracy comparable to the
	refined baselines. This behavior indicates that the coupled-coordinate
	Newton refinement and coupling-aligned delay refinement used in
	CC-NOMP do not compromise spatial-domain estimation accuracy.
	It also indicates that angle estimation is not the main source
	of the RMSE differences among the estimators under the considered fixed
	receive aperture. The main differences among the estimators are
	instead observed in the delay--Doppler dimensions, where the
	AFDM chirp-induced coupling and the associated off-grid ridge
	directly affect the final physical estimates.
	
	Fig.~\ref{fig:main_rmse}(b) compares the normalized delay RMSE.
	Coarse-grid OMP and Standard 3D NOMP exhibit clear high-SNR
	delay error floors under the tested off-grid setting. Fractional
	Local ML reduces part of the delay floor through local fractional
	search, but its accuracy is still affected by the finite search
	window and grid density. In contrast, CC-NOMP consistently
	suppresses the delay error floor and closely follows the deterministic
	delay CRB in the medium- and high-SNR regions. This delay
	improvement is important because the physical Doppler estimate is
	reconstructed through the coupled relation $\nu=\eta+\rho\ell$.
	Hence, residual delay errors can be propagated through the coupling
	factor $\rho$ in the reconstructed Doppler estimate, which makes
	accurate delay refinement critical for suppressing the Doppler
	error floor.
	
	Fig.~\ref{fig:main_rmse}(c) presents the normalized Doppler RMSE.
	Coarse-grid OMP and Standard 3D NOMP show weak SNR dependence
	at moderate and high SNRs, indicating that their Doppler errors are
	dominated by off-grid effects rather than additive noise in this
	SNR region. For Coarse-grid OMP, the floor mainly comes from
	coarse-grid mismatch and hard support selection. For Standard 3D NOMP, the remaining floor is further related
	to coarse initialization bias and the sensitivity of local
	physical-coordinate refinement under strong local delay--Doppler
	coupling. Fractional Local ML lowers this floor through local fractional
	search. At medium and high SNRs, CC-NOMP attains the lowest
	Doppler RMSE among the considered model-matched methods and closely
	follows the deterministic Doppler CRB. These results support the effectiveness of the overall
	coupling-aware refinement structure in reducing the off-grid
	delay--Doppler error floors.
	
	\begin{figure}[!t]
		\centering
		\includegraphics[width=0.88\columnwidth]{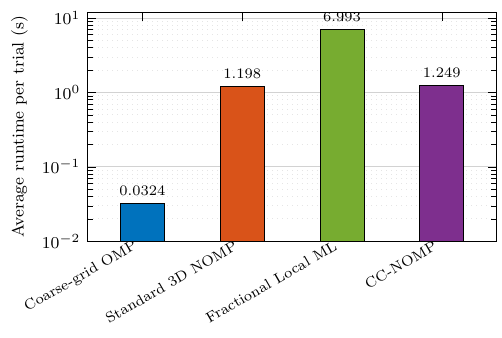}
		\caption{Measured MATLAB runtime per Monte Carlo trial over the considered SNR points.}
		\label{fig:runtime_avg}
	\end{figure}
	
Fig.~\ref{fig:main_rmse}(d) further reports the RMSE of the
auxiliary coupling coordinate $\eta=\nu-\rho\ell$ under the
common $\eta$-aware target assignment. This metric evaluates
estimation accuracy in the auxiliary coordinate associated with
the AFDM-induced coupled ridge. Since $\nu=\eta+\rho\ell$, accurate recovery of the
physical delay and Doppler parameters requires both a small
coupling-coordinate error and accurate localization of the physical
delay--Doppler parameters along the coupled ridge. At medium and high
SNRs, the refined baselines achieve much smaller $\eta$
RMSEs than their physical Doppler RMSEs in
Fig.~\ref{fig:main_rmse}(c), while retaining nonzero delay
error floors in Fig.~\ref{fig:main_rmse}(b), indicating that
they estimate the auxiliary coordinate more accurately than
the corresponding physical delay--Doppler location along the ridge. Nevertheless, they
retain high-SNR $\eta$ error floors, whereas CC-NOMP
continues to improve with SNR and closely follows the
deterministic $\eta$ CRB in the medium- and high-SNR
regions. These results support the proposed design, which
combines coupled-coordinate ridge characterization with
coupling-aligned delay refinement to recover the physical
delay--Doppler parameters.

		\begin{figure}[!t]
		\centering
		\includegraphics[width=0.92\columnwidth]{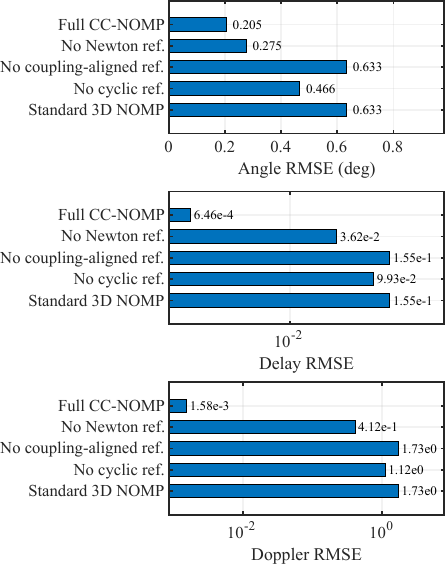}
		\caption{Ablation study of CC-NOMP at SNR = 10 dB.}
		\label{fig:ablation}
	\end{figure}

	Fig.~\ref{fig:runtime_avg} compares the measured MATLAB runtime
	averaged over the considered SNR points and Monte Carlo trials.
	CC-NOMP has a runtime comparable to that of Standard 3D NOMP,
	while remaining substantially faster than Fractional Local ML.
	This agrees with the complexity analysis, since CC-NOMP avoids the dense local fractional search used by
	Fractional Local ML and incorporates coupled-coordinate refinement
	and coupling-aligned delay refinement into the NOMP-type procedure. The measurements were obtained in MATLAB R2024b on a
	Windows 11 Pro desktop with an Intel Core Ultra 7 265K CPU
	and 32 GB RAM. The reported runtimes exclude one-time
	coarse-dictionary construction and fixed waveform-dependent
	preprocessing.

	\subsection{Ablation Study and Target-Number Sensitivity}
	\label{subsec:diagnostic_checks}

Fig.~\ref{fig:ablation} evaluates the major refinement stages in
Section~III-B using 500 Monte Carlo trials at SNR $=10$ dB. Full
CC-NOMP uses the complete procedure. The three ablation variants
disable Newton, coupling-aligned delay, and cyclic refinement,
respectively, while leaving the remaining stages unchanged; for the
No Newton variant, the number of Newton iterations is set to zero.
Standard 3D NOMP serves as the conventional physical-parameter
refinement baseline.

Removing Newton refinement substantially increases the angle, delay,
and Doppler RMSEs, showing that the remaining stages cannot replace
Newton-type continuous localization. Removing coupling-aligned delay
refinement causes the largest delay--Doppler degradation among the
tested ablations, supporting its role in refinement along the
AFDM-induced coupling ridge. Removing cyclic refinement also degrades
performance, consistent with increased residual inter-target
interference. Standard 3D
NOMP performs close to the variant without coupling-aligned refinement
in delay and Doppler, further supporting the contribution of this
refinement. Overall, Fig.~\ref{fig:ablation} confirms the distinct
roles of the major refinement stages in Section~III-B.

	\begin{figure}[!t]
		\centering
		\includegraphics[width=0.96\columnwidth]{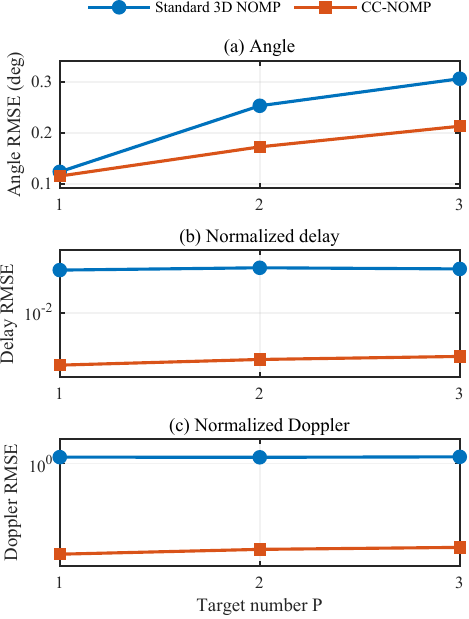}
		\caption{Target-number sensitivity over 500 Monte Carlo trials at SNR = 10 dB: (a) angle, (b) normalized delay, and (c) normalized Doppler.}
		\label{fig:sensitivity_target}
	\end{figure}
	
Fig.~\ref{fig:sensitivity_target} evaluates the target-number
sensitivity for $P=1,2$, and $3$ using 500 Monte Carlo trials.
CC-NOMP maintains lower delay and Doppler RMSEs than Standard
3D NOMP. The angle RMSE increases as the number of targets grows,
because the fixed receive aperture and observation length must
resolve more overlapping target responses. The delay--Doppler
advantage remains visible because cyclic refinement constructs
interference-reduced residuals and applies coupling-aligned delay
refinement to the updated residuals. These results characterize the
tested target-number range and do not represent an exhaustive study
over all possible target configurations.

	\section{Conclusion}
	\label{sec:conclusion}
	
	This paper has studied multi-target angle, continuous normalized
	delay, and normalized Doppler estimation for a colocated
	AFDM-ISAC sensing architecture. A transform-domain sparse sensing model was formulated
	from the fractional DAFT-domain response, where the AFDM chirp
	parameter introduces delay--Doppler coupling and the corresponding
	matching-score surface exhibits a local ridge. Based on this observation, the proposed CC-NOMP
	estimator combines coarse support initialization,
	coupled-coordinate local parameterization, coupling-aligned delay
	refinement, coefficient re-estimation, and cyclic multi-target
	refinement. The auxiliary coordinate provides a structured description of
	the dominant AFDM-induced ridge and supports safeguarded
	coupled-coordinate Newton refinement and Doppler reconstruction. Coupling-aligned
	delay refinement improves estimation of the physical delay--Doppler
	parameters along this ridge, while cyclic refinement suppresses
	residual inter-target interference. The ablation results showed that removing Newton refinement
	degraded estimation accuracy, removing coupling-aligned delay
	refinement caused the largest observed delay--Doppler degradation
	among the tested ablations, and removing cyclic refinement caused
	an additional performance loss. The deterministic CRB and the complexity
	analysis further characterized the local estimation limit and the
	dominant computational cost. Under the considered matched-model setting, CC-NOMP achieves
	lower normalized delay and Doppler error floors than the evaluated
	model-matched baselines while maintaining
	comparable angle-estimation accuracy.

	\appendices
	\begingroup
	\setlength{\abovedisplayskip}{4pt plus 1pt minus 1pt}
	\setlength{\belowdisplayskip}{4pt plus 1pt minus 1pt}
	\setlength{\abovedisplayshortskip}{2pt plus 1pt minus 1pt}
	\setlength{\belowdisplayshortskip}{2pt plus 1pt minus 1pt}
	\section{Derivation of the DAFT-Domain Fractional Delay--Doppler Response}
	\label{app:daft_derivation}
	
	This appendix derives the DAFT-domain fractional delay--Doppler response in \eqref{eq:daft_response}. Starting from the within-symbol received signal in \eqref{eq:rx_periodic}, the conjugate DAFT demodulation kernel at bin $m'$ is
	\begin{equation}
		D_{m'}^{*}(k')
		=
		\frac{1}{\sqrt{N_{\mathrm c}}}
		e^{-\jj2\pi\left(c_1 k'^2 + \frac{m'k'}{N_{\mathrm c}}+c_2m'^2\right)} .
		\label{eq:app_daft_kernel}
	\end{equation}
	Thus,
	\begin{equation}
		y_{n,n_{\mathrm{R}}}(m')
		=
		\sum_{k'=0}^{N_{\mathrm c}-1}
		y_{n,n_{\mathrm{R}}}(k')D_{m'}^{*}(k') .
		\label{eq:app_daft_demod}
	\end{equation}
	Substituting \eqref{eq:rx_periodic} into \eqref{eq:app_daft_demod} yields
	\begin{equation}
		\begin{aligned}
			y_{n,n_{\mathrm{R}}}(m')
			=
			&\sum_{p=1}^{P}
			\alpha_p a_{\mathrm{R}}(n_{\mathrm{R}},\theta_p)
			\Gamma_n(\nu_p)
			\\
			& \times
			\frac{1}{N_{\mathrm c}}
			\sum_{m=0}^{N_{\mathrm c}-1}
			x_{n,m}\\
			&\sum_{k'=0}^{N_{\mathrm c}-1}
			e^{\jj2\pi\Omega_{p,m,m'}(k')}
			+w_{n,n_{\mathrm{R}}}(m'),
		\end{aligned}
		\label{eq:app_substitution}
	\end{equation}
	where
	\begin{equation}
		\begin{aligned}
			\Omega_{p,m,m'}(k')
			&=
			c_1(k'-\ell_p)^2
			+\frac{m(k'-\ell_p)}{N_{\mathrm c}}
			+c_2m^2
			\\
			&\quad
			+\frac{\nu_p k'}{N_{\mathrm c}}
			-c_1k'^2
			-\frac{m'k'}{N_{\mathrm c}}
			-c_2m'^2 .
		\end{aligned}
		\label{eq:app_total_phase}
	\end{equation}
	Collecting the $k'$-independent and $k'$-dependent terms gives
	\begin{equation}
		\begin{aligned}
			\Omega_{p,m,m'}(k')
			&=
			c_1\ell_p^2
			-\frac{m\ell_p}{N_{\mathrm c}}
			+c_2(m^2-m'^2)
			\\
			&\quad
			+
			\frac{
				(m-m')-2N_{\mathrm c} c_1\ell_p+\nu_p
			}{N_{\mathrm c}}k' .
		\end{aligned}
		\label{eq:app_phase_expand}
	\end{equation}
	Define
	\begin{equation}
		\zeta_{p,m,m'}
		=
		(m-m')-2N_{\mathrm c} c_1\ell_p+\nu_p .
		\label{eq:app_zeta_def}
	\end{equation}
	Then the finite geometric sum over $k'$ is
	\begin{equation}
		\begin{aligned}
			\sum_{k'=0}^{N_{\mathrm c}-1}
			e^{\jj(2\pi/N_{\mathrm c})\zeta_{p,m,m'}k'}
			=
			\frac{
				e^{\jj2\pi\zeta_{p,m,m'}}-1
			}{
				e^{\jj(2\pi/N_{\mathrm c})\zeta_{p,m,m'}}-1
			}.
		\end{aligned}
		\label{eq:app_geometric_sum}
	\end{equation}
	Since $\rho=2N_{\mathrm c} c_1$, this expression is consistent with the coupling phase in \eqref{eq:coupling_phase} and the auxiliary coordinate in \eqref{eq:eta_def}.

	\section{Analytical Derivatives of the AFDM Atom}
	\label{app:atom_derivatives}
	
	All derivatives in this appendix are taken with respect to real-valued parameters, while the atom itself is complex-valued; hence the resulting gradient and Hessian are real-valued derivatives of the real matching score. This appendix lists the atom derivatives used by the Newton refinement and CRB evaluation. Write
	\begin{equation}
		\psi_{m,m'}(\ell,\nu)
		=
		E_{m,m'}(\ell)G(\zeta_{m,m'}),
		\label{eq:app_psi_finite_sum}
	\end{equation}
	where
	\begin{equation}
		E_{m,m'}(\ell)
		=
		e^{\jj2\pi[c_1\ell^2-m\ell/N_{\mathrm c}+c_2(m^2-m'^2)]},
		\label{eq:app_E_def}
	\end{equation}
	and
	\begin{equation}
		G(\zeta)=\sum_{q=0}^{N_{\mathrm c}-1} e^{\jj2\pi q\zeta/N_{\mathrm c}}.
		\label{eq:app_G_def}
	\end{equation}
	Its first two derivatives are
	\begin{equation}
		G'(\zeta)
		=
		\sum_{q=0}^{N_{\mathrm c}-1}
		\jj\frac{2\pi q}{N_{\mathrm c}}e^{\jj2\pi q\zeta/N_{\mathrm c}},
		\label{eq:app_G_prime}
	\end{equation}
	\begin{equation}
		G''(\zeta)
		=
		\sum_{q=0}^{N_{\mathrm c}-1}
		\left(\jj\frac{2\pi q}{N_{\mathrm c}}\right)^2
		e^{\jj2\pi q\zeta/N_{\mathrm c}}.
		\label{eq:app_G_second}
	\end{equation}
	Using
	\begin{equation}
		\zeta_{m,m'}=(m-m')-\rho\ell+\nu,\qquad
		\rho=2N_{\mathrm c} c_1,
		\label{eq:app_zeta_generic}
	\end{equation}
	one obtains $\partial\zeta/\partial\ell=-\rho$ and
	$\partial\zeta/\partial\nu=1$. Let
	$\phi_{\ell}=2c_1\ell-m/N_{\mathrm c}$. The first-order derivatives of $\psi$ are
	\begin{equation}
		\frac{\partial \psi}{\partial \ell}
		=
		E\left(\jj2\pi\phi_{\ell}G-\rho G'\right),
		\label{eq:app_psi_l}
	\end{equation}
	\begin{equation}
		\frac{\partial \psi}{\partial \nu}
		=
		EG',
		\label{eq:app_psi_nu}
	\end{equation}
	where $E=E_{m,m'}(\ell)$ and $G$, $G'$, and $G''$ are evaluated at
	$\zeta_{m,m'}$. The second-order derivatives are
	\begin{equation}
		\frac{\partial^2 \psi}{\partial \ell^2}
		=
		E\left[
		(\jj2\pi\phi_{\ell})^2G+\jj4\pi c_1G
		-2\rho\jj2\pi\phi_{\ell}G'
		+\rho^2G''
		\right],
		\label{eq:app_psi_ll}
	\end{equation}
	\begin{equation}
		\frac{\partial^2 \psi}{\partial \ell\partial\nu}
		=
		E\left(\jj2\pi\phi_{\ell}G'-\rho G''\right),
		\label{eq:app_psi_lnu}
	\end{equation}
	\begin{equation}
		\frac{\partial^2 \psi}{\partial \nu^2}
		=
		EG''.
		\label{eq:app_psi_nunu}
	\end{equation}
	The symbol-domain factor is
	\begin{equation}
		b_{n,m'}(\ell,\nu)
		=
		\frac{1}{N_{\mathrm c}}\sum_{m=0}^{N_{\mathrm c}-1}
		x_{n,m}\psi_{m,m'}(\ell,\nu).
		\label{eq:app_b_def}
	\end{equation}
	For $s,t\in\{\ell,\nu\}$, its first- and second-order derivatives are
	\begin{equation}
		\frac{\partial b_{n,m'}}{\partial s}
		=
		\frac{1}{N_{\mathrm c}}\sum_{m=0}^{N_{\mathrm c}-1}
		x_{n,m}
		\frac{\partial\psi_{m,m'}}{\partial s},
		\label{eq:app_b_first_derivative}
	\end{equation}
	and
	\begin{equation}
		\frac{\partial^2 b_{n,m'}}{\partial s\partial t}
		=
		\frac{1}{N_{\mathrm c}}\sum_{m=0}^{N_{\mathrm c}-1}
		x_{n,m}
		\frac{\partial^2\psi_{m,m'}}{\partial s\partial t}.
		\label{eq:app_b_second_derivative}
	\end{equation}
	
	For the receive-array and Doppler-symbol factors, define
	\begin{equation}
		\kappa_{n_{\mathrm{R}}}
		=
		\frac{2\pi d_{\mathrm{R}}(n_{\mathrm{R}}-1)}{\lambda},
		\qquad
		\omega_n=\frac{2\pi t_n}{N_{\mathrm c}}.
		\label{eq:app_kappa_omega}
	\end{equation}
	Then
	\begin{equation}
		\frac{\partial a_{\mathrm{R}}}{\partial\theta}
		=
		\jj\kappa_{n_{\mathrm{R}}}\cos\theta\,a_{\mathrm{R}},
		\label{eq:app_ar_theta}
	\end{equation}
	\begin{equation}
		\frac{\partial^2 a_{\mathrm{R}}}{\partial\theta^2}
		=
		(-\jj\kappa_{n_{\mathrm{R}}}\sin\theta
		-\kappa_{n_{\mathrm{R}}}^2\cos^2\theta)a_{\mathrm{R}},
		\label{eq:app_ar_thetatheta}
	\end{equation}
	and
	\begin{equation}
		\frac{\partial\Gamma_n}{\partial\nu}
		=
		\jj\omega_n\Gamma_n,\qquad
		\frac{\partial^2\Gamma_n}{\partial\nu^2}
		=
		-\omega_n^2\Gamma_n.
		\label{eq:app_gamma_derivatives}
	\end{equation}
	The angle derivatives above are written for a radian-valued angle. If a degree-valued angle is used in implementation, the first- and second-order derivatives with respect to that variable are scaled by $\pi/180$ and $(\pi/180)^2$, respectively.
	The physical-coordinate derivatives of each entry of
	$\mathbf{a}(\theta,\ell,\nu)$ follow from the product rule applied to
	$a_{\mathrm{R}}(n_{\mathrm{R}},\theta)\Gamma_n(\nu)b_{n,m'}(\ell,\nu)$.
	
	Finally, because Newton refinement uses
	$\mathbf{z}=[\theta,\ell,\eta]^T$ with $\nu=\eta+\rho\ell$, the coupled-coordinate first derivatives are
	\begin{equation}
		\frac{\partial \mathbf{a}}{\partial \theta_z}
		=
		\mathbf{a}_{\theta},\quad
		\frac{\partial \mathbf{a}}{\partial \ell_z}
		=
		\mathbf{a}_{\ell}+\rho\mathbf{a}_{\nu},\quad
		\frac{\partial \mathbf{a}}{\partial \eta}
		=
		\mathbf{a}_{\nu}.
		\label{eq:app_chain_first}
	\end{equation}
	The corresponding second-order derivatives include
	\begin{equation}
		\mathbf{a}_{\ell_z\ell_z}
		=
		\mathbf{a}_{\ell\ell}
		+2\rho\mathbf{a}_{\ell\nu}
		+\rho^2\mathbf{a}_{\nu\nu},
		\label{eq:app_chain_ll}
	\end{equation}
	\begin{equation}
		\mathbf{a}_{\ell_z\eta}
		=
		\mathbf{a}_{\ell\nu}+\rho\mathbf{a}_{\nu\nu},
		\qquad
		\mathbf{a}_{\eta\eta}
		=
		\mathbf{a}_{\nu\nu}.
		\label{eq:app_chain_lnu}
	\end{equation}
	The terms involving $\theta_z$ follow the same chain rule, e.g.,
	$\mathbf{a}_{\theta_z\ell_z}=\mathbf{a}_{\theta\ell}+\rho\mathbf{a}_{\theta\nu}$,
	$\mathbf{a}_{\theta_z\eta}=\mathbf{a}_{\theta\nu}$, and
	$\mathbf{a}_{\theta_z\theta_z}=\mathbf{a}_{\theta\theta}$.
	
	The score-gradient and Hessian entries used in \eqref{eq:gradient_hessian}
	are computed from these atom derivatives. Let $x_i$ denote the $i$th
	component of $\mathbf{z}=[\theta,\ell,\eta]^T$, and let
	$\mathbf{a}(\mathbf{z})=\mathbf{a}(\theta,\ell,\eta+\rho\ell)$. The
	normalized matching score is
	\begin{equation}
		S_z(\mathbf z)
		=
		\frac{|\mathbf a^H(\mathbf z)\mathbf r_p|^2}
		{\mathbf a^H(\mathbf z)\mathbf a(\mathbf z)} .
		\label{eq:app_score_definition}
	\end{equation}
	For compactness, define
	\begin{equation}
		h=\mathbf{a}^H(\mathbf{z})\mathbf{r}_p,
		\qquad
		d=\mathbf{a}^H(\mathbf{z})\mathbf{a}(\mathbf{z}).
		\label{eq:app_score_aux_defs}
	\end{equation}
	Let
	\begin{equation}
		h_i=\mathbf{a}_i^H\mathbf{r}_p,
		\qquad
		h_{ij}=\mathbf{a}_{ij}^H\mathbf{r}_p,
		\label{eq:app_score_h_terms}
	\end{equation}
	\begin{equation}
		d_i=2\operatorname{Re}\{\mathbf{a}_i^H\mathbf{a}\},\qquad
		d_{ij}=2\operatorname{Re}\{\mathbf{a}_{ij}^H\mathbf{a}
		+\mathbf{a}_i^H\mathbf{a}_j\},
		\label{eq:app_score_d_terms}
	\end{equation}
	where $\mathbf a_i=\partial\mathbf a(\mathbf z)/\partial x_i$ and
	$\mathbf a_{ij}=\partial^2\mathbf a(\mathbf z)/\partial x_i\partial x_j$.
	Then the gradient entry is
	\begin{equation}
		[g_t]_i
		=
		\frac{
			2\operatorname{Re}\{h^*h_i\}d
			-|h|^2d_i
		}{d^2}.
		\label{eq:app_score_gradient}
	\end{equation}
	The Hessian entry is
	\begin{equation}
		\begin{aligned}
			{}[H_t]_{ij}
			&=
			\frac{2\operatorname{Re}\{h^*h_{ij}+h_i^*h_j\}}{d}
			\\
			&\quad-
			\frac{
				2\operatorname{Re}\{h^*h_i\}d_j
				+2\operatorname{Re}\{h^*h_j\}d_i
				+|h|^2d_{ij}
			}{d^2}
			\\
			&\quad+
			\frac{2|h|^2d_i d_j}{d^3}.
		\end{aligned}
		\label{eq:app_score_hessian}
	\end{equation}

	\endgroup
	\balance
	\bibliographystyle{IEEEtran}
	\bibliography{references_verified_final}

@IEEEtranBSTCTL{IEEEexample:BSTcontrol,
  CTLuse_forced_etal       = "yes",
  CTLmax_names_forced_etal = "6",
  CTLnames_show_etal       = "1"
}

@article{Hassanien2019SPM,
  author  = {A. Hassanien and M. G. Amin and E. Aboutanios and B. Himed},
  title   = {Dual-Function Radar Communication Systems: A Solution to the Spectrum Congestion Problem},
  journal = {IEEE Signal Processing Magazine},
  volume  = {36},
  number  = {5},
  pages   = {115--126},
  month   = sep,
  year    = {2019}
}

@article{Liu2020TCOMM,
  author  = {F. Liu and C. Masouros and A. P. Petropulu and H. Griffiths and L. Hanzo},
  title   = {Joint Radar and Communication Design: Applications, State-of-the-Art, and the Road Ahead},
  journal = {IEEE Transactions on Communications},
  volume  = {68},
  number  = {6},
  pages   = {3834--3862},
  month   = jun,
  year    = {2020}
}

@article{ALiu2022COMST,
  author  = {A. Liu and Z. Huang and M. Li and Y. Wan and W. Li and T. X. Han and C. Liu and R. Du and D. K. P. Tan and J. Lu and Y. Shen and F. Colone and K. Chetty},
  title   = {A Survey on Fundamental Limits of Integrated Sensing and Communication},
  journal = {IEEE Communications Surveys \& Tutorials},
  volume  = {24},
  number  = {2},
  pages   = {994--1034},
  year    = {2022}
}

@article{Zhong2023JointWaveform,
	author  = {K. Zhong and J. Hu and C. Pan and M. Deng and J. Fang},
	title   = {Joint Waveform and Beamforming Design for {RIS}-Aided
	{ISAC} Systems},
	journal = {IEEE Signal Processing Letters},
	volume  = {30},
	pages   = {165--169},
	year    = {2023},
	doi     = {10.1109/LSP.2023.3242554}
}

@article{Ren2026TwoTimescale,
	author  = {Z. Ren and C. Pan and H. Ren and D. Wang and L. Xu
	and J. Wang},
	title   = {Two-Timescale Design for {AP} Mode Selection and
	Power Allocation of Cooperative {ISAC} Networks},
	journal = {IEEE Transactions on Wireless Communications},
	volume  = {25},
	pages   = {815--831},
	year    = {2026},
	doi     = {10.1109/TWC.2025.3587068}
}

@article{Liu2022JSAC,
  author  = {F. Liu and Y. Cui and C. Masouros and J. Xu and T. X. Han and Y. C. Eldar and S. Buzzi},
  title   = {Integrated Sensing and Communications: Toward Dual-Functional Wireless Networks for {6G} and Beyond},
  journal = {IEEE Journal on Selected Areas in Communications},
  volume  = {40},
  number  = {6},
  pages   = {1728--1767},
  month   = jun,
  year    = {2022}
}

@inproceedings{Hadani2017WCNC,
  author    = {R. Hadani and S. Rakib and M. Tsatsanis and A. Monk and A. J. Goldsmith and A. F. Molisch and R. Calderbank},
  title     = {Orthogonal Time Frequency Space Modulation},
  booktitle = {Proc. IEEE Wireless Communications and Networking Conference (WCNC)},
  pages     = {1--6},
  year      = {2017}
}

@article{Mohammed2022BITS,
  author  = {S. K. Mohammed and R. Hadani and A. Chockalingam and R. Calderbank},
  title   = {{OTFS}---A Mathematical Foundation for Communication and Radar Sensing in the Delay--{Doppler} Domain},
  journal = {IEEE BITS the Information Theory Magazine},
  volume  = {2},
  number  = {2},
  pages   = {36--55},
  month   = nov,
  year    = {2022}
}

@article{Gaudio2020TWC,
  author  = {L. Gaudio and M. Kobayashi and G. Caire and G. Colavolpe},
  title   = {On the Effectiveness of {OTFS} for Joint Radar Parameter Estimation and Communication},
  journal = {IEEE Transactions on Wireless Communications},
  volume  = {19},
  number  = {9},
  pages   = {5951--5965},
  month   = sep,
  year    = {2020}
}

@article{Yuan2024MWC,
  author  = {W. Yuan and L. Zhou and S. K. Dehkordi and S. Li and P. Fan and G. Caire and H. V. Poor},
  title   = {From {OTFS} to {DD-ISAC}: Integrating Sensing and Communications in the Delay {Doppler} Domain},
  journal = {IEEE Wireless Communications},
  volume  = {31},
  number  = {6},
  pages   = {152--160},
  month   = dec,
  year    = {2024}
}

@article{Zhang2023OTFSRadarSensing,
  author  = {K. Zhang and Z. Li and W. Yuan and Y. Cai and F. Gao},
  title   = {Radar Sensing via {OTFS} Signaling},
  journal = {China Communications},
  volume  = {20},
  number  = {9},
  pages   = {34--45},
  month   = sep,
  year    = {2023}
}

@article{Rou2024SPM,
  author  = {H. S. Rou and G. T. F. de Abreu and J. Choi and D. {Gonz{\'a}lez G.} and M. Kountouris and Y. L. Guan and O. Gonsa},
  title   = {From Orthogonal Time--Frequency Space to Affine Frequency-Division Multiplexing: A Comparative Study of Next-Generation Waveforms for Integrated Sensing and Communications in Doubly Dispersive Channels},
  journal = {IEEE Signal Processing Magazine},
  volume  = {41},
  number  = {5},
  pages   = {71--86},
  month   = sep,
  year    = {2024},
  doi     = {10.1109/MSP.2024.3422653}
}

@article{Bemani2023TWC,
  author  = {A. Bemani and N. Ksairi and M. Kountouris},
  title   = {Affine Frequency Division Multiplexing for Next Generation Wireless Communications},
  journal = {IEEE Transactions on Wireless Communications},
  volume  = {22},
  number  = {11},
  pages   = {8214--8229},
  month   = nov,
  year    = {2023}
}

@inproceedings{Benzine2023GLOBECOM,
  author    = {W. Benzine and A. Bemani and N. Ksairi and D. T. M. Slock},
  title     = {Affine Frequency Division Multiplexing for Communications on Sparse Time-Varying Channels},
  booktitle = {Proc. IEEE Global Communications Conference (GLOBECOM)},
  pages     = {4921--4926},
  year      = {2023}
}

@article{Yin2025MWC,
  author  = {H. Yin and Y. Tang and A. Bemani and M. Kountouris and Y. Zhou and X. Zhang and Y. Liu and G. Chen and K. Yang and F. Liu and C. Masouros and S. Li and G. Caire and P. Xiao},
  title   = {Affine Frequency Division Multiplexing: Extending {OFDM} for Scenario-Flexibility and Resilience},
  journal = {IEEE Wireless Communications},
  volume  = {32},
  number  = {6},
  pages   = {200--208},
  month   = dec,
  year    = {2025}
}

@article{Rou2026AFDM6G,
  author  = {H. S. Rou and K. R. R. Ranasinghe and V. Savaux and G. T. F. de Abreu and D. {Gonz{\'a}lez G.} and C. Masouros},
  title   = {Affine Frequency Division Multiplexing ({AFDM}) for {6G}: Properties, Features, and Challenges},
  journal = {IEEE Communications Standards Magazine},
  volume  = {10},
  number  = {2},
  pages   = {216--225},
  month   = jun,
  year    = {2026},
  doi     = {10.1109/MCOMSTD.2025.3643183}
}

@inproceedings{Bedeer2025AFDMAmbiguity,
  author    = {E. Bedeer},
  title     = {Ambiguity Function Analysis of Affine Frequency Division Multiplexing for {ISAC}},
  booktitle = {Proc. IEEE Global Communications Conference (GLOBECOM)},
  pages     = {2850--2855},
  month     = dec,
  year      = {2025}
}

@article{Yin2026JSACAmbiguity,
  author  = {H. Yin and Y. Tang and Y. Ni and Z. Wang and G. Chen and J. Xiong and K. Yang and M. Kountouris and Y. L. Guan and Y. Zeng},
  title   = {Ambiguity Function Analysis of {AFDM} Signals for Integrated Sensing and Communications},
  journal = {IEEE Journal on Selected Areas in Communications},
  volume  = {44},
  pages   = {196--211},
  year    = {2026},
  doi     = {10.1109/JSAC.2025.3611936}
}

@article{Ni2026PulseShapedAF,
  author  = {Y. Ni and F. Liu and H. Yin and Y. Tang and Y. Ma and Z. Wang},
  title   = {Ambiguity Function Analysis of {AFDM} Under Pulse-Shaped Random {ISAC} Signaling},
  journal = {IEEE Transactions on Wireless Communications},
  volume  = {25},
  pages   = {13619--13635},
  year    = {2026},
  doi     = {10.1109/TWC.2026.3670065}
}

@inproceedings{Ni2022ISWCS,
  author    = {Y. Ni and Z. Wang and P. Yuan and Q. Huang},
  title     = {An {AFDM}-Based Integrated Sensing and Communications},
  booktitle = {Proc. International Symposium on Wireless Communication Systems (ISWCS)},
  pages     = {1--6},
  year      = {2022},
  doi       = {10.1109/ISWCS56560.2022.9940346}
}

@article{Bemani2024WCL,
  author  = {A. Bemani and N. Ksairi and M. Kountouris},
  title   = {Integrated Sensing and Communications With Affine Frequency Division Multiplexing},
  journal = {IEEE Wireless Communications Letters},
  volume  = {13},
  number  = {5},
  pages   = {1255--1259},
  month   = may,
  year    = {2024}
}

@article{Ni2025TWC,
  author  = {Y. Ni and P. Yuan and Q. Huang and F. Liu and Z. Wang},
  title   = {An Integrated Sensing and Communications System Based on Affine Frequency Division Multiplexing},
  journal = {IEEE Transactions on Wireless Communications},
  volume  = {24},
  number  = {5},
  pages   = {3763--3779},
  month   = may,
  year    = {2025},
  doi     = {10.1109/TWC.2025.3532993}
}

@inproceedings{Temiz2025SPAWC,
  author    = {M. Temiz and C. Masouros},
  title     = {Affine Frequency Division Multiplexing With Subcarrier Power-Level Index Modulation for Integrated Sensing and Communications},
  booktitle = {Proc. IEEE International Workshop on Signal Processing Advances in Wireless Communications (SPAWC)},
  pages     = {1--5},
  year      = {2025}
}

@article{Zhang2025AFDMISACFramework,
  author  = {F. Zhang and Z. Wang and T. Mao and T. Jiao and Y. Zhuo and M. Wen and W. Xiang and S. Chen and G. K. Karagiannidis},
  title   = {{AFDM}-Enabled Integrated Sensing and Communication: Theoretical Framework and Pilot Design},
  journal = {IEEE Journal on Selected Areas in Communications},
  volume  = {44},
  pages   = {310--324},
  year    = {2026}
}

@article{Zhu2024WCL,
  author  = {J. Zhu and Y. Tang and F. Liu and X. Zhang and H. Yin and Y. Zhou},
  title   = {{AFDM}-Based Bistatic Integrated Sensing and Communication in Static Scatterer Environments},
  journal = {IEEE Wireless Communications Letters},
  volume  = {13},
  number  = {8},
  pages   = {2245--2249},
  month   = aug,
  year    = {2024}
}

@inproceedings{Luo2025Reconstruction,
  author    = {Y. Luo and Y. Ding and Z. Lin and S. Xue and C. Fan and D. Kong and L. Zhang and S. Deng},
  title     = {{AFDM}-Enabled Integrated Sensing and Communication for {4D} Reconstruction},
  booktitle = {Proc. International Conference on Intelligent Communications and Computing (ICICC)},
  pages     = {211--216},
  year      = {2025}
}

@inproceedings{Chen2025MUSIC,
  author    = {S. Chen and X. Ren and X. Lin and Z. Li and W. Li and L. Mei},
  title     = {High-Accuracy Perception Algorithm Based on {AFDM-ISAC} System},
  booktitle = {Proc. IEEE Vehicular Technology Conference (VTC-Fall)},
  pages     = {1--6},
  year      = {2025},
  doi       = {10.1109/VTC2025-Fall65116.2025.11309964}
}

@article{Li2026DeepLearningAFDM,
	author  = {J. Li and H. Wei and X. Liu and Y. Liu
	and M. Peng},
	title   = {Deep Learning-Enabled {AFDM} Receiver for
	Multi-Target Super-Resolution Sensing in
	High-Mobility {ISAC} Systems},
	journal = {IEEE Transactions on Wireless Communications},
	volume  = {25},
	pages   = {19564--19578},
	year    = {2026},
	doi     = {10.1109/TWC.2026.3709663}
}

@inproceedings{Lu2025DAFTIC,
  author    = {X. Lu and H. Yin and C. Yi and Y. Zhou and J. Zhu and Y. Tang and W. Li and S. Ge},
  title     = {{DAFT}-Domain Interference Cancellation Scheme for Full-Duplex {AFDM ISAC} System},
  booktitle = {Proc. IEEE International Conference on Communications Workshops (ICC Workshops)},
  pages     = {1067--1072},
  year      = {2025},
  doi       = {10.1109/ICCWorkshops67674.2025.11162304}
}

@article{Xiao2026TCOMM,
  author  = {F. Xiao and Z. Li and D. T. M. Slock},
  title   = {Multipath Component Power Delay Profile Based Joint Range and {Doppler} Estimation for {AFDM-ISAC} Systems},
  journal = {IEEE Transactions on Communications},
  volume  = {74},
  pages   = {7993--8007},
  year    = {2026},
  doi     = {10.1109/TCOMM.2026.3686728}
}

@article{Ranasinghe2025TWC,
  author  = {K. R. R. Ranasinghe and H. S. Rou and G. T. F. de Abreu and T. Takahashi and K. Ito},
  title   = {Joint Channel, Data, and Radar Parameter Estimation for {AFDM} Systems in Doubly-Dispersive Channels},
  journal = {IEEE Transactions on Wireless Communications},
  volume  = {24},
  number  = {2},
  pages   = {1602--1619},
  month   = feb,
  year    = {2025}
}

@inproceedings{Ranasinghe2025Blind,
  author    = {K. R. R. Ranasinghe and K. Ando and H. S. Rou and G. T. F. de Abreu and A. Bathelt},
  title     = {Blind Bistatic Radar Parameter Estimation in Doubly-Dispersive Channels},
  booktitle = {Proc. IEEE Wireless Communications and Networking Conference (WCNC)},
  pages     = {1--6},
  year      = {2025}
}

@article{Luo2025IoTJ,
  author  = {Y. Luo and Y. L. Guan and Y. Ge and D. {Gonz{\'a}lez G.} and C. Yuen},
  title   = {A Novel Angle--Delay--{Doppler} Estimation Scheme for {AFDM-ISAC} System in Mixed Near-Field and Far-Field Scenarios},
  journal = {IEEE Internet of Things Journal},
  volume  = {12},
  number  = {13},
  pages   = {22669--22682},
  month   = jul,
  year    = {2025}
}

@article{Zhu2026WCL,
  author  = {S. Zhu and L. Mei and Z. Li and Z. Du and Y. Song},
  title   = {Low Complexity {AFDM}-Based {ISAC} Utilizing Fractional Delay--{Doppler} Feature Extraction},
  journal = {IEEE Wireless Communications Letters},
  volume  = {15},
  pages   = {335--339},
  year    = {2026},
  doi     = {10.1109/LWC.2025.3626429}
}

@inproceedings{Luo2025SBL,
  author    = {Y. Luo and Y. L. Guan and Y. Ge and C. Yuen},
  title     = {Target Sensing With Off-Grid Sparse {Bayesian} Learning for {AFDM-ISAC} System},
  booktitle = {Proc. IEEE International Conference on Communications Workshops (ICC Workshops)},
  pages     = {881--886},
  year      = {2025}
}

@article{Zhu2026FMCWAFDM,
  author  = {J. Zhu and Y. Tang and C. Yi and H. Yin and Y. Ni and F. Liu and Z. Wei and H. Arslan},
  title   = {{ISAC} With Affine Frequency Division Multiplexing: An {FMCW}-Based Signal Processing Perspective},
  journal = {IEEE Transactions on Wireless Communications},
  volume  = {25},
  pages   = {19725--19739},
  year    = {2026},
  doi     = {10.1109/TWC.2026.3706989}
}

@article{Chi2011BasisMismatch,
  author  = {Y. Chi and L. L. Scharf and A. Pezeshki and R. Calderbank},
  title   = {Sensitivity to Basis Mismatch in Compressed Sensing},
  journal = {IEEE Transactions on Signal Processing},
  volume  = {59},
  number  = {5},
  pages   = {2182--2195},
  month   = may,
  year    = {2011},
  doi     = {10.1109/TSP.2011.2112650}
}

@article{Tropp2007TIT,
  author  = {J. A. Tropp and A. C. Gilbert},
  title   = {Signal Recovery From Random Measurements via Orthogonal Matching Pursuit},
  journal = {IEEE Transactions on Information Theory},
  volume  = {53},
  number  = {12},
  pages   = {4655--4666},
  month   = dec,
  year    = {2007},
  doi     = {10.1109/TIT.2007.909108}
}

@article{Ramasamy2016TSP,
  author  = {D. Ramasamy and B. Mamandipoor and U. Madhow},
  title   = {Newtonized Orthogonal Matching Pursuit: Frequency Estimation Over the Continuum},
  journal = {IEEE Transactions on Signal Processing},
  volume  = {64},
  number  = {19},
  pages   = {5066--5081},
  month   = oct,
  year    = {2016}
}

@article{Shah2025NOMPOFDM,
	author  = {S. N. H. Shah and S. Semper and A. U. Khan
	and C. Schneider and J. Robert},
	title   = {Newtonized Orthogonal Matching Pursuit for
	High-Resolution Target Detection in Sparse
	{OFDM} {ISAC} Systems},
	journal = {IEEE Transactions on Vehicular Technology},
	volume  = {74},
	number  = {10},
	pages   = {16137--16151},
	month   = oct,
	year    = {2025},
	doi     = {10.1109/TVT.2025.3573165}
}
	
\end{document}